\documentclass[longauth]{aa}  
\usepackage{graphicx}
\usepackage{txfonts}
\usepackage[colorlinks=true,citecolor=blue]{hyperref}

\usepackage{multirow}
\usepackage{xcolor}
\usepackage{caption}
\usepackage{subcaption}
\usepackage{threeparttable}
\usepackage{rotating}
\usepackage{rotfloat}
\usepackage{float}
\usepackage{moresize}
\usepackage{wrapfig}
\usepackage{multicol}
\usepackage{wrapfig}
\usepackage[normalem]{ulem}
\usepackage{graphicx}
\usepackage{txfonts}
\usepackage[T1]{fontenc}
 \usepackage{graphicx}  
\usepackage{amsmath}    
 \usepackage{color}
\usepackage{amssymb}
\usepackage{epstopdf}
\usepackage{longtable}
\usepackage{natbib}
\usepackage[export]{adjustbox}
\usepackage{tabularx}
\usepackage{subcaption}
\bibpunct{(}{)}{;}{a}{}{,} 
\usepackage{url}

\newcommand{\kms}{km\,s$^{-1}$}
\newcommand{\ms}{m~s$^{-1}$}

\newcommand{\logrhk}{$\rm log\,R^{\prime}_\mathrm{HK}$}

\begin{document}

\title{TOI-3862\,b: A dense super-Neptune deep in the hot Neptune desert}

   \author{Ilaria Carleo\inst{\ref{inafoato},\ref{iac}}
          \and Amadeo Castro-Gonz{\'a}lez\inst{\ref{unige}}
          \and Enric Pall{\'e}\inst{\ref{iac}, \ref{ull}}
          \and Felipe Murgas\inst{\ref{iac}, \ref{ull}}
          \and Grzegorz Nowak \inst{\ref{umk}}
          \and Gaia Lacedelli\inst{\ref{iac}, \ref{ull}}
          \and Thomas Masseron \inst{\ref{iac}, \ref{ull}}
          \and Emily W. Wong  \inst{\ref{unige}}
          \and Patrick Eggenberger  \inst{\ref{unige}}
          \and Vincent Bourrier  \inst{\ref{unige}}
          \and Dawid Jankowski \inst{\ref{umk}}
          \and Krzysztof Go\'zdziewski \inst{\ref{umk}}
          \and Douglas R. Alves \inst{\ref{uchile}, \ref{cata}}
          \and James S. Jenkins \inst{\ref{cata}, \ref{udp}}
          \and Sergio Messina \inst{\ref{inaf-catania}}
          \and Keivan G.\ Stassun \inst{\ref{vanderbilt}}
          \and Jose I. Vines \inst{\ref{uchile}}
          \and Matteo Brogi \inst{\ref{unito}, \ref{inafoato}}
          \and David~R.~Ciardi \inst{\ref{IPAC}}
          \and Catherine A. Clark \inst{\ref{IPAC}}
          \and William Cochran \inst{\ref{utexas}}
          \and Karen A.\ Collins \inst{\ref{harvard}}
          \and Hans J.~Deeg \inst{\ref{iac}, \ref{ull}}
          \and Elise Furlan \inst{\ref{IPAC}}
          \and Davide Gandolfi\inst{\ref{unito}}
          \and Samuel Geraldía González \inst{\ref{iac}, \ref{ull}}
          \and Artie P. Hatzes \inst{\ref{tautenburg}}
          \and Coel Hellier \inst{\ref{keele}}
          \and Steve~B.~Howell \inst{\ref{ames}}
          \and Judith Korth \inst{\ref{unige}}
          \and Jorge Lillo-Box \inst{\ref{cab}}
          \and John H. Livingston \inst{\ref{Tokyo}, \ref{NRAO}, \ref{Sokendai}}
          \and Jaume Orell-Miquel \inst{\ref{utexas}}
          \and Carina M. Persson \inst{\ref{OSO}}
          \and Seth Redfield \inst{\ref{Wesleyan}}
          \and Boris Safonov \inst{\ref{Moscow}}
          \and David Baker \inst{\ref{AustinCollege}}
          \and Rafael Delfin Barrena Delgado \inst{\ref{iac}, \ref{ull}}
          \and Allyson Bieryla \inst{\ref{harvard}}
          \and Andrew Boyle \inst{\ref{IPAC}, \ref{unc}}
          \and Pau Bosch-Cabot \inst{\ref{Girona}}
          \and Núria Casasayas Barris \inst{\ref{iac}, \ref{ull}}
          \and Stavros Chairetas \inst{\ref{iac}}
          \and Jerome P. de Leon \inst{\ref{UniTokyo}}
          \and Izuru Fukuda \inst{\ref{UniTokyo-mds}}
          \and Akihiko Fukui \inst{\ref{UniTokyo}, \ref{iac}}
          \and Pere Guerra \inst{\ref{Girona}}
          \and Kai Ikuta \inst{\ref{Hitotsubashi}}
          \and Kiyoe Kawauchi \inst{\ref{Ritsumeikan}}
          \and Emil Knudstrup \inst{\ref{Aarhus}, \ref{OSO}}
          \and Florence Libotte \inst{\ref{iac}, \ref{sabadell}, \ref{RoyalBelgian}}
          \and Michael B.~Lund \inst{\ref{IPAC}}
          \and Rafael Luque \inst{\ref{Andalucia}}
          \and Eduardo Lorenzo Martín Guerrero de Escalante \inst{\ref{iac}}
          \and Bob Massey \inst{\ref{Villa}}
          \and Edward J. Michaels \inst{\ref{Waffelow}}
          \and Giuseppe Morello \inst{\ref{Andalucia}, \ref{inaf-Palermo}}
          \and Norio Narita \inst{\ref{UniTokyo}, \ref{Tokyo}, \ref{iac}}
          \and Hannu Parvianien \inst{\ref{iac}, \ref{ull}}
          \and Richard P. Schwarz \inst{\ref{harvard}}
          \and Avi Shporer \inst{\ref{Kavli}}
          \and Monika Stangret \inst{\ref{iac}, \ref{ull}}
          \and Noriharu Watanabe \inst{\ref{UniTokyo-mds}}
          \and Cristilyn N. Watkins \inst{\ref{Bozeman}}
          }

   \institute{INAF -- Osservatorio Astrofisico di Torino, Via Osservatorio 20, I-10025, Pino Torinese, Italy\label{inafoato}\\ 
              \email{ilaria.carleo@inaf.it}
        \and Instituto de Astrof\'{i}sica de Canarias (IAC), 38205 La Laguna, Tenerife, Spain \label{iac} 
         \and Astronomical Observatory, University of Geneva, Chemin Pegasi 51b, CH-1290 Versoix, Switzerland \label{unige}  
         \and Departamento de Astrof\'isica, Universidad de La Laguna (ULL), E-38206 La Laguna, Tenerife, Spain \label{ull} 
         \and Institute of Astronomy, Faculty of Physics, Astronomy and Informatics, Nicolaus Copernicus University, Grudziądzka 5, 87-100 Toruń, Poland \label{umk} 
         \and Departamento de Astronomía, Universidad de Chile, Casilla 36-D, 7591245, Santiago, Chile \label{uchile}
         \and Centro de Astrofísica y Tecnologías Afines (CATA), Casilla 36-D, 7591245, Santiago, Chile \label{cata} 
         \and Núcleo de Astronomía, Facultad de Ingeniería y Ciencias, Universidad Diego Portales, Av. Ejército 441 Santiago, Chile \label{udp}  
         \and INAF - Osservatorio Astrofisico di Catania, Via S. Sofia 78, I-95123 Catania, Italy \label{inaf-catania} 
         \and Department of Physics \& Astronomy, Vanderbilt University, Nashville, TN, USA \label{vanderbilt} 
         \and Dipartimento di Fisica, Universit{\'a} degli Studi di Torino, via Pietro Giuria 1, I-10125, Torino, Italy \label{unito} 
         \and NASA Exoplanet Science Institute, IPAC, California Institute of Technology, Pasadena, CA 91125 USA \label{IPAC} 
         \and Center for Astrophysics \textbar \ Harvard \& Smithsonian, 60 Garden Street, Cambridge, MA 02138, USA \label{harvard} 
         \and Th\"uringer Landessternwarte Sternwarte 5 D-07778, Tautenburg, Germany \label{tautenburg} 
         \and Astrophysics Group, Keele University, Staffs ST5 5BG, U.K. \label{keele} 
         \and NASA Ames Research Center, Moffett Field, CA 94035, USA \label{ames} 
         \and Centro de Astrobiolog\'ia (CAB, CSIC-INTA), Depto. de Astrof\'isica, ESAC campus, 28692, Villanueva de la Ca\~nada (Madrid), Spain \label{cab} 
         \and Astrobiology Center, 2-21-1 Osawa, Mitaka, Tokyo 181-8588, Japan \label{Tokyo} 
         \and National Astronomical Observatory of Japan, 2-21-1 Osawa, Mitaka, Tokyo 181-8588,  Japan.\label{NRAO} 
         \and Department of Astronomy, The Graduate University for Advanced Studies (SOKENDAI), 2-21-1 Osawa, Mitaka, Tokyo, Japan \label{Sokendai}  
         \and The University of Texas at Austin, 2515 Speedway, Stop C1402, Austin, Texas 78712-1206 \label{utexas} 
         \and Chalmers University of Technology, Department of Space, Earth and Environment, Onsala Space Observatory, SE-439 92 Onsala, Sweden \label{OSO} 
         \and Astronomy Department and Van Vleck Observatory, Wesleyan University, Middletown, CT 06459, USA \label{Wesleyan} 
         \and Sternberg Astronomical Institute, Lomonosov Moscow State University, 119992 Universitetskii prospekt 13, Moscow, Russia \label{Moscow} 
         \and Physics Department, Austin College, Sherman, TX 75090, USA \label{AustinCollege} 
         \and Department of Physics and Astronomy, University of North Carolina at Chapel Hill, Chapel Hill, NC 27599, USA \label{unc} 
         \and Observatori Astronòmic Albanyà, Camí de Bassegoda S/N, Albanyà 17733, Girona, Spain \label{Girona} 
         \and Komaba Institute for Science, The University of Tokyo, 3-8-1 Komaba, Meguro, Tokyo 153-8902, Japan \label{UniTokyo} 
         \and Department of Multi-Disciplinary Sciences, Graduate School of Arts and Sciences, The University of Tokyo, 3-8-1 Komaba, Meguro, Tokyo 153-8902, Japan \label{UniTokyo-mds} 
         \and Graduate School of Social Data Science, Hitotsubashi University, 2-1 Naka, Kunitachi, Tokyo 186-8601, Japan \label{Hitotsubashi} 
         \and Department of Physical Sciences, Ritsumeikan University, Kusatsu, Shiga 525-8577, Japan \label{Ritsumeikan}  
         \and Stellar Astrophysics Centre, Department of Physics and Astronomy, Aarhus University, Ny Munkegade 120, 8000 Aarhus C, Denmark \label{Aarhus} 
         \and Sabadell Astronomical Society, 08206 Sabadell, Barcelona, Spain \label{sabadell}  
         \and Europlanet Society, Department of Planetary Atmospheres of the Royal Belgian Institute for Space Aeronomy, B-1180 Brussels, Belgium \label{RoyalBelgian} 
         \and Instituto de Astrofísica de Andalucía (IAA-CSIC), Glorieta de la Astronomía s/n, 18008 Granada, Spain  \label{Andalucia} 
         \and Villa '39 Observatory, Landers, CA 92285, USA \label{Villa} 
         \and Waffelow Creek Observatory, 10780 FM 1878, Nacogdoches, TX 75961, USA \label{Waffelow} 
         \and INAF - Osservatorio Astronomico di Palermo, Piazza del Parlamento, 1, 90134 Palermo, Italy. \label{inaf-Palermo}  
         \and Department of Physics and Kavli Institute for Astrophysics and Space Research, Massachusetts Institute of Technology, Cambridge, MA 02139, USA \label{Kavli} 
         \and Bozeman, MT 59718, USA \label{Bozeman} 
             }

\date{Received Date Month YYYY; accepted Date Month YYYY}

 
  \abstract
   {The structure and evolution of close-in exoplanets are shaped by atmospheric loss and migration processes, which give rise to key population features such as the hot Neptune desert, ridge, and savanna — regions of the period-radius space whose boundaries offer critical insights into planetary formation and survival.}
   {As part of the KESPRINT collaboration, we selected the TESS transiting planet candidate TOI-3862.01 for radial velocity follow-up to confirm its planetary nature and characterize its mass and bulk properties. This planet candidate is of particular interest due to its position in the middle of the hot Neptune desert, making it a valuable probe for testing theories of planet migration and atmospheric loss.}
   {We confirmed the planetary nature and determined the mass of TOI-3862.01 (hereinafter TOI-3862\,b) by performing a joint fit with both transit and radial velocity data, precisely characterizing the bulk properties of this planet.}
   {TOI-3862\,b is a super-Neptune on a 1.56-day orbit around a Sun-like star with an effective temperature of 5300$\pm$50\,K. It has a mass of 53.7$_{-2.9}^{+2.8}$ M$_{\oplus}$ and a radius of 5.53$\pm$0.18 R$_{\oplus}$, corresponding to a density of 1.7$\pm$0.2 g/cm³. This places it among the rare population of hot and dense super-Neptune desert planets.}
   {TOI-3862\,b, residing deep in the hot Neptune desert, represents a rare occurrence in an otherwise sparsely populated region, offering a valuable opportunity to probe the processes that may allow planets to survive in such environments.}


\titlerunning{TOI-3862}
\authorrunning{Carleo et al.}

   \maketitle
%

\section{Introduction} 
\label{sec:intro}
Intermediate-size exoplanets, particularly those with radii between 4 and 10~R$_{\oplus}$, exhibit notable distribution features that challenge our understanding of planet formation and evolution. Among these is the so-called hot Neptune desert -- a region of parameter space at short orbital periods ($\lesssim$\,3.2~days) where planets between the sizes of Neptune and Saturn are rare despite their detection advantages \citep[e.g.,][]{2007A&A...461.1185L,Benitez2011,SzaboKiss2011,Mazehetal2016,CastroGonzalez2024}. The origin of this desert remains an open question, with leading theories invoking atmospheric photoevaporation, tidal disruption, and inhibited formation or migration pathways as possible explanations \citep[e.g.,][]{ Mazehetal2016,2016ApJ...820L...8M,2018MNRAS.479.5012O}.

Recent studies have revealed additional planet occurrence structures in the close-in exo-Neptunian landscape. The Neptunian ridge, for instance, is a concentration of Neptune-sized planets at orbital periods of $\sim$3.2–5.7~days \citep{CastroGonzalez2024}, situated between the desert and the more populated Neptunian savanna \citep{2023A&A...669A..63B}. The orbital range of the ridge aligns with that of the well-known hot Jupiter pile-up \citep[e.g.,][]{1999ApJ...526..890C,2003A&A...407..369U}, suggesting that similar evolutionary processes might be affecting close-in giant planets from Neptune to Jupiter sizes. Interestingly, planets located within the desert and ridge regions often exhibit higher densities than their counterparts in the savanna, with a dividing line at $\sim$1~g\,cm$^{-3}$, possibly indicating more extreme atmospheric loss or distinct formation histories \citep{CastroGonzalez2024b}. These features suggest that the transition between planet classes in this regime is driven by a combination of photoevaporation, core composition, and migration dynamics.

In this work, we report the discovery and characterization of TOI-3862\,b, a super-Neptune located squarely within the hot Neptune desert. Its position in this sparsely populated region makes it a valuable target for testing atmospheric loss theories and for probing the formation and survival of intermediate-size planets in extreme environments. We describe the photometric and spectroscopic observations in Section~\ref{sec:obs}, detail the stellar characterization in Section~\ref{sec: Stellar modelling}, present the frequency analysis in Section~\ref{sec:freq}, and describe the joint modeling of the system in Section~\ref{sec:planet}. We also discuss the implications of this discovery with internal structure and atmospheric evolution modeling in Section~\ref{sec-toi-3862-atm} and put the discovered system in context within the wider population in Section \ref{sec:mass-radius}, with our conclusions summarized in Section~\ref{sec:concl}.

\section{Observations} \label{sec:obs}
\subsection{Photometric data}

\subsubsection{\textit{TESS} photometry}
\label{sec:TESS_phot}
TOI-3862 (TIC\,141205978, stellar properties in Table~\ref{tab:star_param}) was photometrically monitored at a 2-minute cadence by the Transiting Exoplanet Survey Satellite (TESS, \citealt{2015JATIS...1a4003R}) between August 2019 and March 2024. The information on the TESS sectors, CCDs (Charge-Coupled Devices), and cameras are reported in Table~\ref{table:tess_obs}.
The Science Processing Operations Center (SPOC) conducted a transit search using an adaptive, noise-compensating matched filter \citep{Jenkins02,Jenkins10,Jenkins2020kepler}, which resulted in the identification of a threshold crossing event (TCE). An initial limb-darkened transit model was fit to the signal \citep{Li:DVmodelFit2019}, and a comprehensive suite of diagnostic tests was applied to evaluate the planetary nature of the event \citep{Twicken:DVdiagnostics2018}. The transit signature was independently detected in full-frame image (FFI) data by the Quick Look Pipeline (QLP) at the Massachusetts Institute of Technology \citep[MIT;][]{Huang2020QLP1,Huang2020QLP2}. Vetting reports were reviewed by the TESS Science Office (TSO), and the signal was subjected to additional scrutiny across multiple sectors. The transit event was consistently recovered in subsequent observations and successfully passed all diagnostic metrics presented in the data validation reports.  

The TESS data were reduced by both the MIT QLP and the TESS SPOC pipeline \citep{jenkins2016}, performing simple aperture photometry \citep[SAP;][]{Twicken10} to produce time series light curves. An additional step to correct for instrumental systematics was performed by the Presearch Data Conditioning (PDCSAP) algorithm \citep{Smith2012_PDCSAP,Stumpe2012}. For our target, we downloaded the SPOC light curve from the Mikulski Archive for Space Telescopes (MAST\footnote{\url{https://mast.stsci.edu/}}) and used the PDCSAP data for the light curve fit (Sect.~\ref{sec:planet}).

\begin{figure*}
\centering
\includegraphics[width=0.43\linewidth,trim=10 10 10 0,clip]{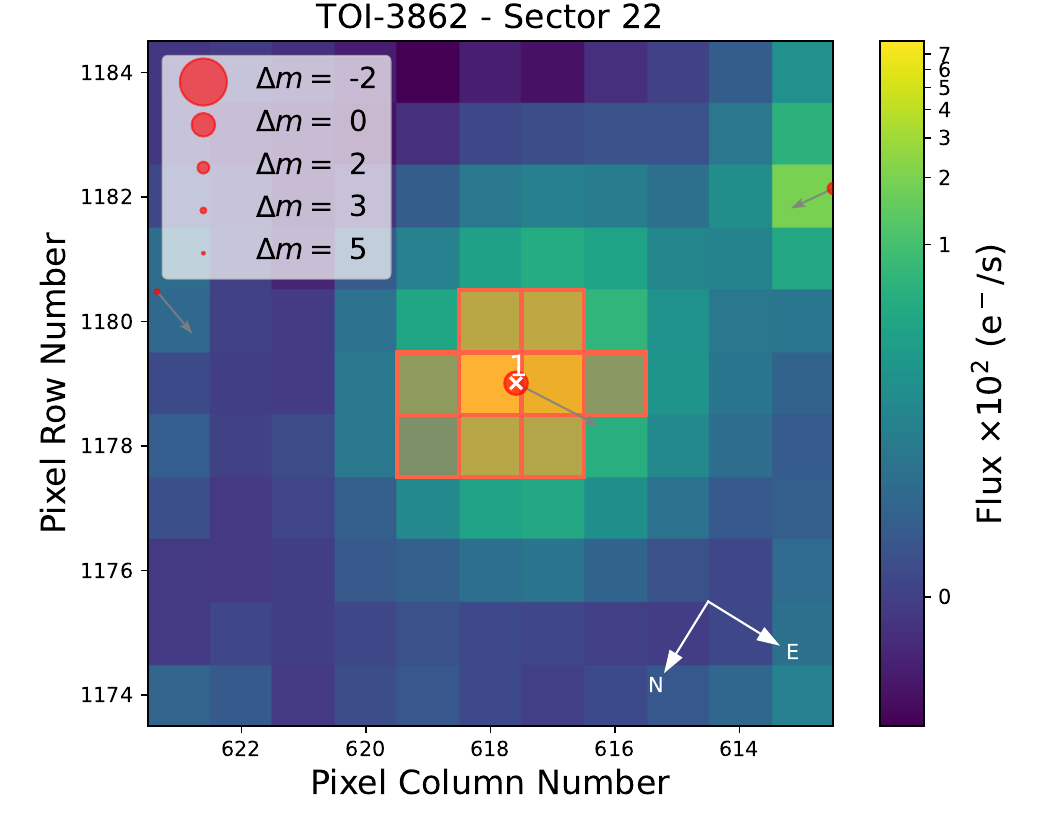}
\includegraphics[width=0.43\linewidth,trim=0 0 0 0,clip]{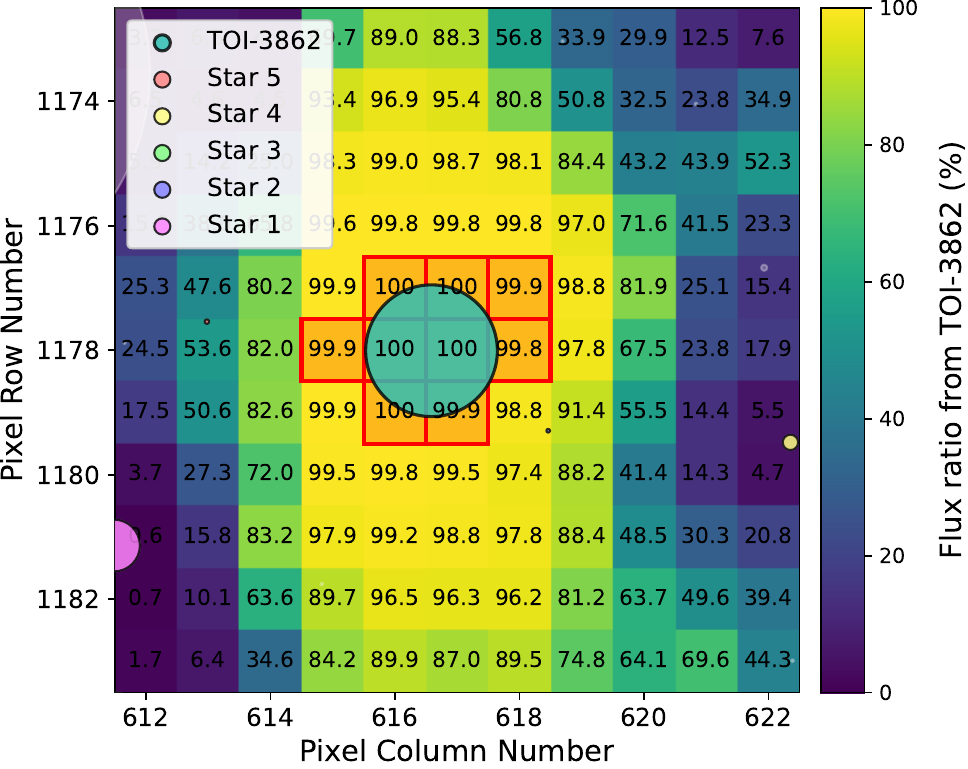}
\caption{\textit{Left}: TESS TPF of Sector 22 for TOI-3862. The color bar represents the electron counts for each pixel. The orange squares denote the pixels chosen by the TESS pipeline for aperture photometry. All sources from Gaia DR3 are overlaid on the plot and depicted as circles of varying sizes, corresponding to their G-mag difference relative to the target (as detailed in the legend). This visualization was generated using the {\tt tpfplotter} code \citep{aller2020}. Gray arrows indicate the proper motion directions for all sources shown in the plot. \textit{Right}: TESS heat maps, generated through \texttt{TESS-cont} \citep{CastroGonzalez2024b}, showing the percentage of the flux in each pixel that comes from the target star. The five most contaminating Gaia DR3 sources are overlaid with sizes scaling with their emitted fluxes.}\label{fig:tpfplotter}
\end{figure*}

We checked for additional sources that could contaminate the TESS flux of our target. To do so, we first used \texttt{tpfplotter}\footnote{\url{https://github.com/jlillo/tpfplotter}} \citep{aller2020} to overlay the sources from Gaia Data Release 3 \citep{2023A&A...674A...1G} onto the target pixel files (TPFs; see Figure~\ref{fig:tpfplotter}), and then used \texttt{TESS-cont}\footnote{\url{https://github.com/castro-gzlz/TESS-cont}} \citep{CastroGonzalez2024b} to compute the flux fractions in the SPOC apertures coming from these sources. For TOI-3862, the most contaminating source is TIC\,141205984 (pixel row 1182, pixel column 613) with a 0.03$\%$ contribution. Following \citet{2018AJ....156..277L,2018AJ....156...78L,2021MNRAS.508..195D,2022MNRAS.509.1075C}, and considering the transit depths of the planetary candidate estimated by the SPOC pipeline, we find that the transits observed in the TOI-3862 aperture cannot be generated by any Gaia contaminant source.

\begin{table}[!htb]
   \caption[]{Stellar properties of TOI-3862.}
     \label{tab:star_param}
     \small
     \centering
       \begin{tabular}{lcl}
         \hline \hline
         \noalign{\smallskip}
         Parameter   &    & Ref \\
         \noalign{\smallskip}
         \hline
         \noalign{\smallskip}
$\alpha$ (ICRS, 2016.0)              & 12:23:36.14  & Gaia DR3\\
$\delta$ (ICRS, 2016.0)             &  +50:32:40.65 & Gaia DR3\\
$\mu_{\alpha}$ (mas/yr)     & 41.308  & Gaia DR3\\
$\mu_{\delta}$ (mas/yr)     & -11.254  & Gaia DR3\\
RV     (km\,s$^{-1}$)       & -27.841$\pm$0.483  & Gaia DR3  \\
$\pi$  (mas)                & 4.0060  & Gaia DR3\\
Distance (pc)     & 245.7778   & Gaia DR3\\
\noalign{\medskip}
$V$ (mag)                    &  12.351$\pm$0.046     & TIC v1.8\\
$B$ (mag)                 & 13.064$\pm$0.1  & TIC v1.8 \\
$G$ (mag)                    &  12.2326$\pm$0.0002  & Gaia DR3 \\
$G_{\rm BP}$-$G_{\rm RP}$ (mag)      &  0.968 &  Gaia DR3\\
TESS (mag)              & 11.7201$\pm$0.007   & TIC v1.8\\
J$_{\rm 2MASS}$ (mag)   &  11.030$\pm$0.023  & TIC v1.8 \\
H$_{\rm 2MASS}$ (mag)     &  10.639$\pm$0.022  & TIC v1.8 \\
K$_{\rm 2MASS}$ (mag)   &  10.548$\pm$0.016  & TIC v1.8 \\
\noalign{\medskip}
$S_{\rm MW}$             & 0.16$\pm$0.01  & This work \\
$\log R^{'}_{\rm HK}$    & -5.05$\pm$0.10  &  This work \\ 
\noalign{\medskip}

         \hline
      \end{tabular}
\end{table}

\begin{table}
\caption{Summary of TESS observations for TOI-3862, including sector, CCD number, camera, and observation cadence.}
\centering
\begin{tabular}{cccc}
\hline\hline
Sector  & CCD Number  & Camera       & Cadence [min]  \\[1mm]
\hline
 15 &  4  & 4 & 2\\
       22 &    2& 2 & 2\\
       48 &  2  & 1 & 2\\
       49 &  2  & 2 & 2\\
       75 &  1  & 1 & 2\\
       76 &  2  & 1 & 2\\[1mm]
\hline
\end{tabular}

\label{table:tess_obs}
\end{table}

\subsubsection{TFOP follow-up light curves}
\label{sec:TFOP_lc}

To check the deblended TESS transit depths, place limits on transit depth chromaticity, and refine the transit ephemerides, we acquired ground-based follow-up photometry of the fields around TOI-3862 as part of the \textit{TESS} Follow-up Observing Program \citep[TFOP;][]{collins:2019}\footnote{\url{https://tess.mit.edu/followup}}. We used the {\tt TESS Transit Finder}, which is a customized version of the {\tt Tapir} software package \citep{Jensen:2013}, to schedule our transit observations. 

We observed three transit windows of the planet candidate using KeplerCam on the 1.2\,m telescope at the Fred Lawrence Whipple Observatory, the 0.4\,m telescope at Observatori Astron\`{o}mic Albany\`{a} (OAA) located in Albany\`{a}, Girona, Spain, the LCOGT 0.35\,m network node at Haleakala Observatory on Maui, Hawai'i (LCO 2m0 Hal), USA, and the multiband imager MuSCAT2 \citep{Narita2019}, mounted on the 1.5 m Telescopio Carlos Sánchez (TCS) at the Teide Observatory in Tenerife, Spain. MuSCAT2 is equipped with four CCDs, enabling simultaneous imaging in the $g'$, $r'$, $i'$, and $z_s$ bands with minimal readout time. Each CCD has $1024 \times 1024$ pixels, providing a field of view of $7.4 \times 7.4$ arcmin$^2$. To prevent saturation, the telescope was slightly defocused. Due to connection issues, the $i'$-band camera was unavailable during the observations. Exposure times were set to 15, 45, and 45 seconds in the $g'$, $r'$, and $z_s$ bands, respectively. The raw images were processed using the MuSCAT2 data reduction pipeline \citep{Parviainen2019}, which performs dark and flat-field corrections, aperture photometry, and transit model fitting, accounting for instrumental systematics.

All LCOGT images were calibrated by the standard LCOGT {\tt BANZAI} pipeline \citep{McCully:2018}, and differential photometric data were extracted using {\tt AstroImageJ} \citep{Collins:2017}. {\tt AstroImageJ} was used to both calibrate and extract differential photometric data from images provided by all other observatories. All follow-up light curves were extracted using small circular photometric apertures that excluded all flux from the nearest known Gaia DR3 catalog neighbor of the target star. The light curves are available on the {\tt EXOFOP-TESS} website\footnote{\url{https://exofop.ipac.caltech.edu/tess/}} and summarized in Table \ref{table:carleo-SG1-phot-obs}, and were jointly fit in the manner described in Sect.~\ref{sec:planet}.

\begin{table*}
\caption{Summary of TFOP ground-based light curve follow-up of TOI-3862.}
\centering
\begin{tabular}{lllccl}
\hline\hline
Telescope & Date  & Filter  &  Phot. Aper.      & Nearest Gaia DR3  & Transit  \\
          & [UTC] &         &  Radius [arcsec]  & neighbor [arcsec] & Coverage \\
\hline
KeplerCam 1.2\,m & 2022-02-05 & Sloan $i'$       & 6.7  &  47.9  & full \\
OAA 0.4\,m       & 2022-05-09 & $\mathrm{I_c}$   & 10.1 &  47.9  & full \\
TCS-Muscat2 1.52\,m   &  2022-03-09    &   g,r,z    &  10.9      &  47.9  &  full  \\
LCO-Hal 0.35\,m  & 2023-03-31 & Sloan $g'$       & 6.3  &  47.9  & full \\[2mm]
\hline
\end{tabular}

\label{table:carleo-SG1-phot-obs}
\end{table*}

\subsubsection{Ground-based archival data}

We analyzed  publicly available archival data of TOI-3862 from ASAS-SN \citep{Shappee:2014,Kochanek2017,Hart:2023}, the WASP transit survey \citep{2006PASP..118.1407P}, and ZTF \citep{Bellm:2019,Masci:2019} to look for the rotational modulations of the host star.
The ASAS-SN $g$ band time series has 470 points with a baseline of $\sim$2200 days, while the $V$ band observations consist of 260 points spanning $\sim$2000 days. The ZTF $g$ band dataset has 660 observations with a baseline of $\sim$1950 days. No significant peak was found in the periodogram for the ASAS-SN and ZTF datasets. WASP covered this target in 2011 and 2012, though spanning only 60 nights in each year, with a total of 17\,000 data points. Again, no rotational modulation is seen, with a 95\% confidence upper limit of 2 mmag.

\subsection{Spectroscopic data}
\label{sec:spectroscopic_data}

TOI-3862 is part of the follow-up carried out within the {\tt KESPRINT} collaboration\footnote{\url{www.kesprint.science}.} in order to determine the mass of small and intermediate-size planets, and has been observed with the visible spectrograph HARPS-N \citep{Cosentinoetal2014} at Telescopio Nazionale Galileo (TNG) in La Palma, Spain, through the observing programs CAT19A\_162, CAT21A\_119, CAT22A\_111 (PI: Nowak), ITP19\_1 (PI: Palle), CAT20B\_80 (PI: Casasayas), CAT23A\_52, and CAT23B\_74 (PI: Carleo).

We collected 28 HARPS-N radial velocities (RVs) between 25 March 2022 and 26 June 2023 UT with an exposure time of 3300~s, and average S/N of 35. We used a G2 mask,  appropriate for the spectral type of the star, and a CCF width of 30~km\,s$^{-1}$, obtaining  an average RV uncertainty of 3~m\,s$^{-1}$ and a RMS (root mean square) of 24\,m\,s$^{-1}$. The data for TOI-3862 are listed in Table \ref{tab:rvdata_toi3862}. We also obtained the \logrhk, S-index, bisector, CCF contrast, and CCF full width at half maximum (FWHM) from the HARPS-N DRS, while the chromospheric index CRX, differential line width (dLW), H-alpha, and the sodium lines Na$_1$ and Na$_2$ were obtained from \texttt{serval} \citep{Zechmeister2018}.

\subsection{High-resolution imaging}
As part of the validation process for transiting exoplanets, we assessed the possible contamination by nearby stars. This is also important for the derived planetary radii \citep{ciardi2015,FurlanHowell2017,FurlanHowell2020, lillo-box12,lillo-box14b,lillo-box24}. For this, our star was observed with optical speckle, lucky-imaging, and near-infrared adaptive optics imaging.

	\subsubsection{Optical speckle imaging}

TOI-3862 was observed on 31 March 2022 UT with the speckle polarimeter on the 2.5-m telescope at the Caucasian Observatory of Sternberg Astronomical Institute (SAI) of Lomonosov Moscow State University. An electron-multiplying CCD detector Andor iXon 897 was used \citep{Safonov2017}. We used the $I_\mathrm{c}$ band with an angular resolution of 0.083$^{\prime\prime}$.  No companions were detected. The detection limits at distances of $0.25$ and $1.0^{\prime\prime}$ from the star are $\Delta I_\mathrm{c}=3.8^m$ and $6.9^m$.

\subsubsection{Near-infrared AO imaging}
Observations of TOI-3862 were made on 19 May 2022 UT with the PHARO instrument \citep{hayward2001} on the Palomar Hale (5m) behind the P3K natural guide star adaptive optics (AO) system \citep{dekany2013}. The pixel scale for PHARO is $0.025\arcsec$. The data were collected in a standard 5-point quincunx dither pattern. The reduced science frames were combined into a single mosaiced image with a final resolution of $\sim$0.1\arcsec. 
	
The sensitivity of the final combined AO images was determined by injecting simulated sources azimuthally around the primary target every $20^\circ $ at separations of integer multiples of the central source's FWHM \citep{furlan2017}. The brightness of each injected source was scaled until standard aperture photometry detected it with a $5\sigma $ significance. The final $5\sigma$ limit at each separation was determined from the average of all of the determined limits at that separation and the uncertainty on the limit was set by the RMS dispersion of the azimuthal slices at a given radial distance.

\section{Stellar modeling} \label{sec: Stellar modelling}

We employed three methods to retrieve the stellar parameters. Below, we describe each method in detail. 

\subsection{\texttt{BACCHUS}+PARAM}\label{subsec:thomas} 
The analysis of the stellar spectra was carried out by using the \texttt{BACCHUS} code (\citealp{2016ascl.soft05004M}, with updates from \citealp{2022ApJS..262...34H}), relying on the MARCS model atmospheres \citep{2008A&A...486..951G}, and using the co-added HARPS-N spectrum. Effective temperatures (T$_{\rm eff}$) were derived by requiring no trend of the \ion{Fe}{I} lines abundances against their respective excitation potential. Surface gravities ($\log g$) were determined by requiring an ionization balance between \ion{Fe}{I} lines and the \ion{Fe}{II} line. Microturbulence velocity values ($\xi_t$) were also derived by requiring no trend of Fe line abundances against their equivalent widths. The output metallicity is represented by the average abundance of the \ion{Fe}{I} lines.
We used the HARPS-N spectra to measure the stellar projected rotational velocity ($v \sin i$) using the average of the Fe line broadening after having subtracted the instrument and natural broadening (we neglect macroturbulence velocity for such a cool dwarf). This technique led to only an upper limit of $<$3.5 km~s$\rm ^{-1}$, suggesting a relatively long stellar rotation period ($\textgreater$13 days).

In a second step, we used the Bayesian tool PARAM \citep{2012MNRAS.427..127B,2017MNRAS.467.1433R} to derive the stellar mass and radius, utilizing the updated \textit{Gaia} luminosity along with our spectroscopic temperature. However, such Bayesian tools underestimate the error budget as they do not take into account the systematic errors between different sets of isochrones due to the various underlying assumptions in the respective stellar evolutionary codes (e.g., convection treatment, boundary conditions, opacities, and element diffusion). In order to have an estimate of those systematic errors, we combined the results of the two sets of isochrones provided by PARAM (i.e., MESA and Parsec) and added the difference between the two sets of results to the error budget provided by PARAM. We emphasize that although the use of the two sets of isochrones may mitigate the underlying systematic errors, our formal error budget for radius and luminosity may still be underestimated, as is demonstrated by \citet{Tayar2022}. However, it appears that the results of this error analysis do not significantly affect our analysis of TOI-3862. The results of this analysis are shown in Table \ref{Table: stellar spectroscopic parameters}.

\subsection{\texttt{SPECIES}+ARIADNE} 
We also derived the stellar parameters through the \texttt{SPECIES} algorithm \citep{SotoJenkins2018} applied to the co-added spectrum to derive the chemical abundances. These abundances then act as priors for ARIADNE, which is a Bayesian model averaging (BMA) algorithm for nearby stars \citep{VinesJenkins2022} to get constraints on the bulk parameters. 

\texttt{SPECIES} estimates key atmospheric parameters, including T$_{\rm eff}$, metallicity ([Fe/H]), $\log g$, and $\xi_t$. Initially, the code calculates the equivalent widths ($W$) of Fe I and Fe II lines. These measurements, along with an interpolated grid of ATLAS9 atmospheric models \citep{castelli2004new}, are provided to MOOG \citep{sneden1973nitrogen}, which solves the radiative transfer equation. During this process, Fe line abundances are analyzed as functions of excitation potential and $W$, under the assumption of local thermodynamic equilibrium. Atmospheric parameters are iteratively refined until no correlation is found between the iron abundance and the excitation potential or reduced equivalent width ($W/\lambda$). Finally, the $v \sin i$ is calculated using temperature-based calibrators and by fitting the observed coadded spectral absorption lines with synthetic line profiles.
The spectroscopic results are then analyzed with ARIADNE, which leverages several stellar evolution models within a BMA framework to derive posterior distributions of fundamental stellar properties, including radius and mass, while systematically marginalizing over model-dependent uncertainties (the obtained values are displayed in Table \ref{Table: stellar spectroscopic parameters}).

\subsection{Spectral energy distribution}\label{sec:sed} 

\begin{figure}
\centering
\includegraphics[width=0.95\linewidth,trim=80 70 50 50,clip]{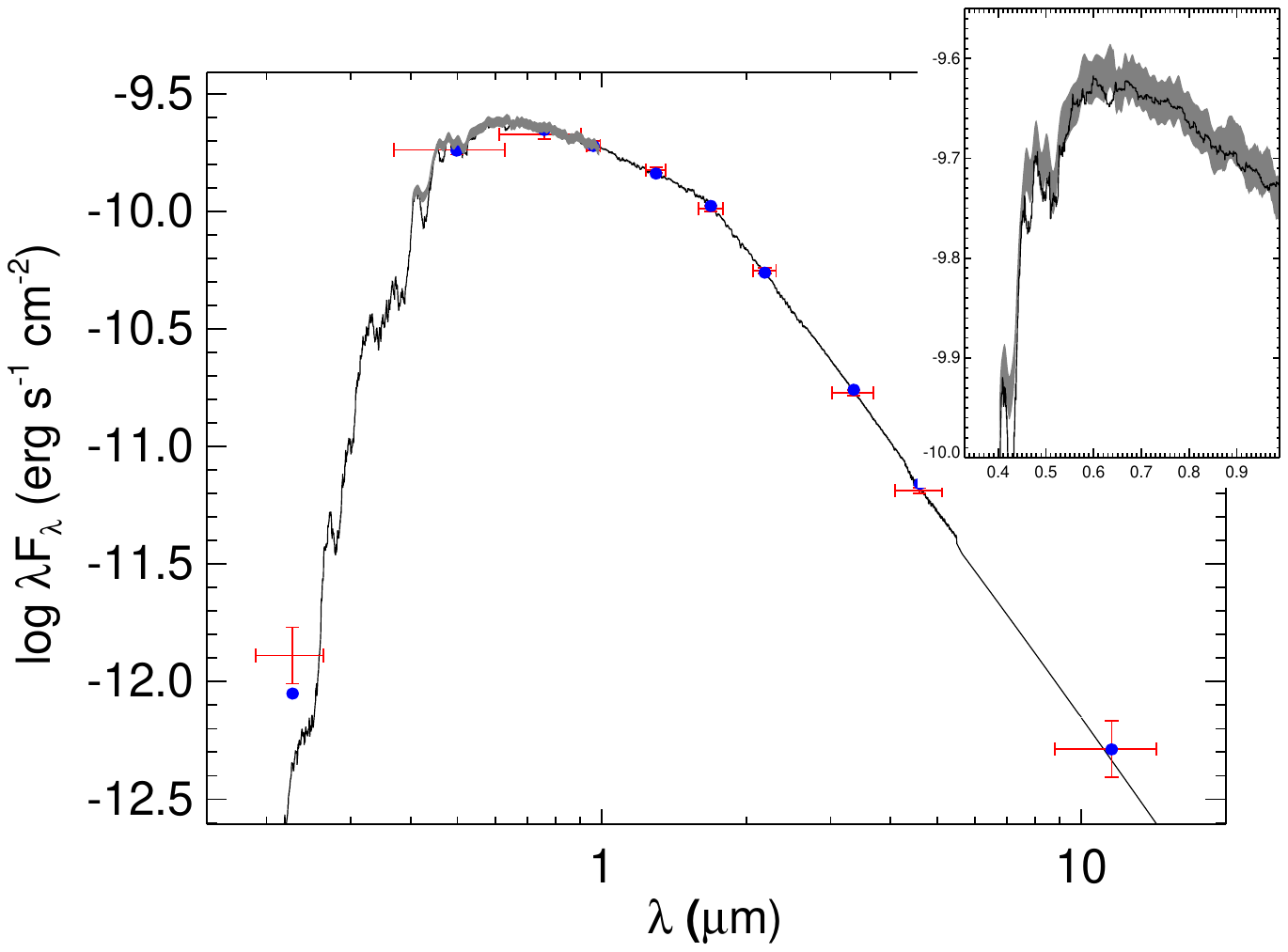}
\caption{Spectral energy distributions of TOI-3862. Red symbols represent the observed photometric measurements, where the horizontal bars represent the effective width of the pass-band. Blue symbols are the model fluxes from the best-fit PHOENIX atmosphere model (black). The absolute flux-calibrated {\it Gaia\/} spectrum is shown as a gray swathe in the inset figure. \label{fig:sed}}
\end{figure}

An analysis of the broadband spectral energy distribution (SED) of the star was performed together with the {\it Gaia\/} DR3 parallax, in order to determine an empirical measurement of the stellar radius \citep{Stassun:2016,Stassun:2017,Stassun:2018}. Where available, the $JHK_S$ magnitudes were sourced from {\it 2MASS}, the W1--W4 magnitudes from {\it WISE}, the $G_{\rm BP}$ and $G_{\rm RP}$ magnitudes from {\it Gaia}, and the NUV magnitude from {\it GALEX}. The absolute flux-calibrated {\it Gaia\/} spectrum was also utilized. Together, the available photometry spans the full stellar SED over the wavelength range of at least 0.4--10~$\mu$m and as much as 0.2--20~$\mu$m (Figure~\ref{fig:sed}). 

A fit using PHOENIX stellar atmosphere models \citep{Husser:2013} was performed, adopting  $T_{\rm eff}$, [Fe/H], and $\log g$ from the spectroscopic analysis. The extinction, $A_V$, was fit for, limited to the maximum line-of-sight value from the Galactic dust maps of \citet{Schlegel:1998}. 
Integrating the (unreddened) model SED gives the bolometric flux at Earth, $F_{\rm bol}$. 
Taking the $F_{\rm bol}$ together with the {\it Gaia\/} parallax directly gives the bolometric luminosity, $L_{\rm bol}$. 
The Stefan-Boltzmann relation then gives the stellar radius, $R_\star$. 
In addition, the stellar mass was estimated using the empirical relations of \citet{Torres:2010}. 
Finally, the system age may be estimated from the observed chromospheric activity, $R'_{\rm HK}$, and the empirical activity-age relations of \citet{Mamajek:2008}; these same relations also predict the stellar rotation period via an empirical gyrochronology relation.

The best fit has $A_V\,=\,0.02\,\pm\,0.02$, with a reduced $\chi^2$ of 1.5, $F_{\rm bol}\,=\,3.1991\,\pm\,0.0050\,\times\,10^{-10}$~erg~s$^{-1}$~cm$^{-2}$, $L_{\rm bol}\,=\,0.6214\,\pm\,0.0021$~L$_\odot$, and estimated rotation period of $22\,\pm\,3$~d, the latter from the empirical activity-rotation relations of \citet{Mamajek:2008}. The resulting fit is shown in Figure~\ref{fig:sed} and the derived parameters are reported in Table \ref{Table: stellar spectroscopic parameters}. The metallicity and age estimates confirm that TOI-3862 is part of the Galactic thin disk population.

\begin{table*}
\centering
 \caption{Spectroscopic  parameters for TOI-3862, as derived in this work. }   
\begin{tabular}{llccccccc }
\hline \hline 
     \noalign{\smallskip} \noalign{\smallskip}
Method  & $T_\mathrm{eff}$  & $\log g_\star$ & [Fe/H]   & Mass & Radius &   Age & $v$ $\sin$i \\  
& (K)  &(cgs)& (dex) & (M$_{\odot}$)& (R$_{\odot}$)& (Gyr)  & \kms   \\
\noalign{\smallskip}
 \hline
     \noalign{\smallskip} \noalign{\smallskip}
\texttt{BACCHUS}+PARAM$^a$  & 5300$\pm$50  &  4.4$\pm$0.1  & 0.11$\pm$0.09    & 0.90$\pm$0.07  & 0.92$\pm$0.03  & \ldots   & \textless\, 3.5 \\

SPECIES+ARIADNE   &  5380$\pm$50    &  4.3$\pm$0.1   &     0.09$\pm$0.01  &\ldots  & \ldots & 4.6$_{-4.1}^{+2.8}$  & 2.3 $\pm$ 0.3\\ 

SED &   \ldots  &\ldots   &\ldots  & 0.94$\pm$0.06 & 0.94$\pm$0.02 & 7.5$\pm$1.9  & \ldots \\ \noalign{\smallskip}
\hline 
\end{tabular}  
\label{Table: stellar spectroscopic parameters}
\tablefoot{
\tablefoottext{a}{Values adopted for the joint fit in  Sect.~\ref{sec:planet}.}
}
\end{table*}

\section{Periodogram analysis} \label{sec:freq}
\begin{figure}
     \includegraphics[width=0.49\textwidth]{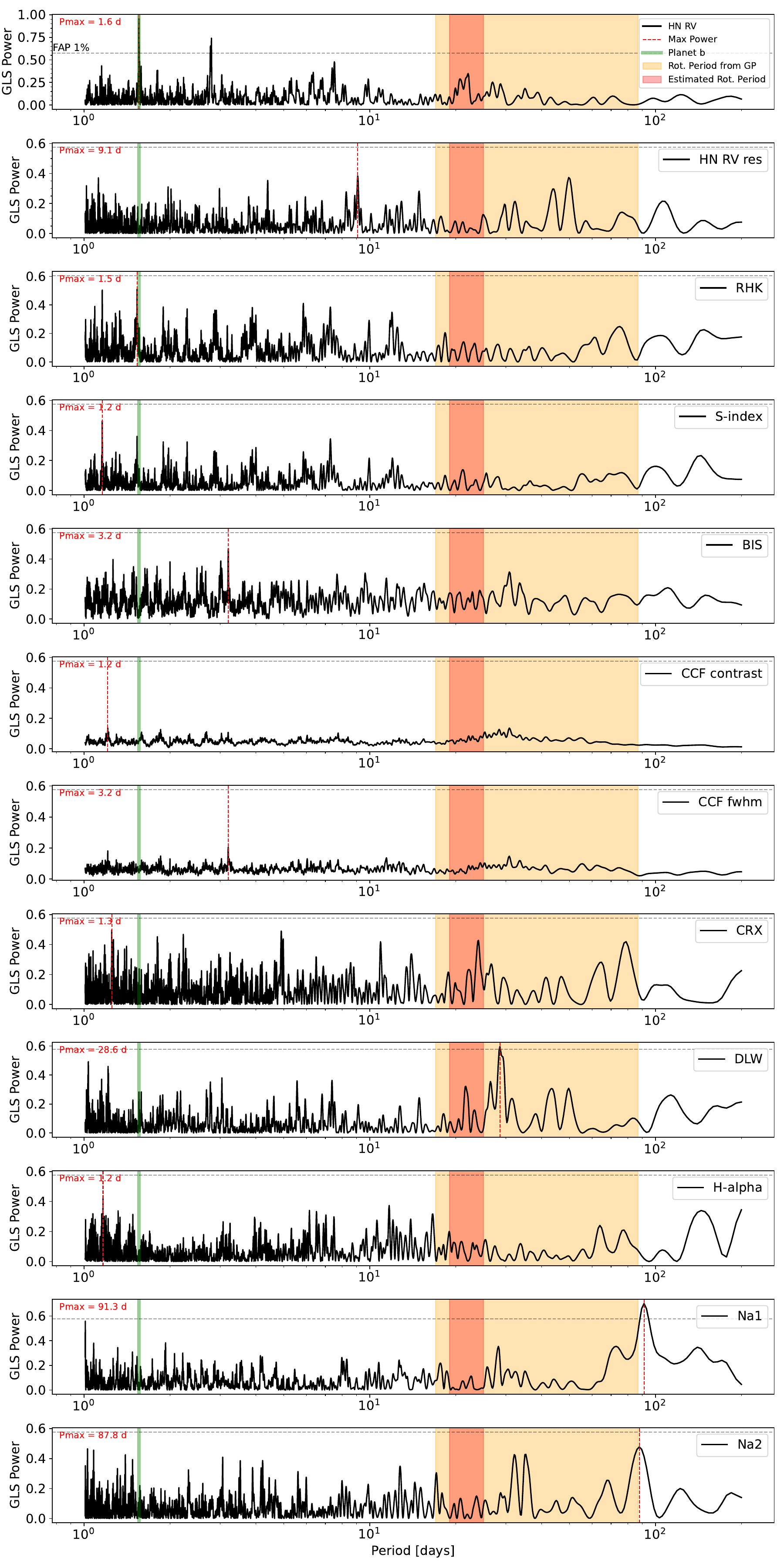}
    \caption{GLS periodograms for TOI-3862. The periodograms were performed for the RV dataset, its residuals (after subtracting the fit model), and all the activity indicators obtained through the HARPS-N DRS (\logrhk, S-index, Bisector, CCF Contrast, and CCF FWHM) and \texttt{serval} (chromospheric index CRX, dLw, H-alpha, and the sodium lines Na$_1$ and Na$_2$). The stellar rotation period, highlighted by the orange region in the periodograms, corresponds to the result of the joint fit using GPs with broad period boundaries (2–100 days), as described in Section \ref{sec:planet}. The red region indicates the rotation period estimated from the SED analysis (see Section \ref{sec:sed}). The dashed red line marks the period corresponding to the maximum power in the periodograms, while the green line denotes the orbital period of planets b.}
    \label{fig:periodogram3862}
\end{figure}

We performed a frequency analysis by employing the GLS periodograms in order to investigate the signals in the TESS light curves and RV dataset (Table \ref{tab:rvdata_toi3862}), as well as in the activity indicators (Tables \ref{tab:drsdata_toi3862} and \ref{tab:servaldata_toi3862}). While the TESS data do not show any significant additional periodicity, the RV periodograms show the most significant peak at the period of the transiting planet (see Figure~\ref{fig:periodogram3862}). The frequency analysis reveals several peaks in the periodograms of the stellar activity indicators, with varying levels of significance. We note that the maximum power period of the \logrhk\ index coincides with the periodicity of the TESS transit signal (i.e. $\simeq$1.5 days). Interestingly, stellar activity synchronized with the orbit of close-in planets is typically interpreted as a manifestation of magnetic star-planet interactions \citep[MSPIs, e.g.,][]{2000ApJ...533L.151C}, and, today, dozens of these signs have been reported in systems with massive close-in giant planets \citep[e.g.,][]{2003ApJ...597.1092S,2005ApJ...622.1075S,2019NatAs...3.1128C,2024A&A...684A.160C}. TOI-3862, hosting a massive super-Neptune (see Sect.~\ref{sec:planet}) well inside the Alfvén radius ($\simeq$10~$\rm R_{\odot}$), represents an interesting candidate for the detection and study of MSPIs \citep[e.g., see the  <$\rm MADK$> $-$ $\rm M_{\rm p} sin (i) / P_{\rm orb}$ correlation by][]{2018haex.bookE..20S}. However, given that the \logrhk\ signal is still not significant (the peak is below FAP=1$\%$), we remain cautious about its interpretation. Overall, since each activity index displays different periodic signals, the Gaussian process (GP) model used in the joint fit (Section \ref{subsec:joint}) was assigned with wide, uninformative priors on the stellar rotation period (ranging from 2 to 100 days) to allow for flexibility in capturing the relevant timescales. In the periodogram plots (Figure~\ref{fig:periodogram3862}), two shaded regions highlight the estimated stellar rotation periods: the red region corresponds to the value inferred from the SED analysis (Section~\ref{sec:sed}), while the orange region (49$_{-32}^{+38}$ days) shows the posterior distribution from the joint fit using the GP (Section~\ref{subsec:joint}).

The rotational period of TOI-3862 inferred from the GP encompasses the SED-based estimate and some weak peaks in the activity periodograms fall within this range. However, due to the low significance of these signals (and the fact that this star is chromospherically inactive (i.e., \logrhk $\textless$ -4.75; \citealt{VaughanPreston1980, Middelkoop1982, Henryetal1996, Gondoin2020}), no firm conclusion can be drawn regarding the stellar rotation.

\section{Planetary system modeling}\label{sec:planet}

\label{subsec:joint}

We performed a global fit with RVs and photometric transit data in order to obtain the planetary system parameters. We used the package {\tt PyORBIT}\footnote{Available at \url{https://github.com/LucaMalavolta/PyORBIT}} (\citealt{Malavoltaetal2016,Malavoltaetal2018}). For the transit modeling of the light curves, we used the package \texttt{batman} \citep{batman}, which implements the quadratic limb darkened transit model by \citet{2002ApJ...580L.171M}, and an additional local polynomial trend is included for each transit (since the TESS light curves were not previously detrended). This model includes the time of the first inferior conjunction, $T_c$, the orbital period, $P$, the impact parameter, $b$, the eccentricity, $e$, and argument of periastron, $\omega$, following the parametrization from \citealt{Eastman2013} ($\sqrt{e}\cos\omega$,$\sqrt{e}\sin\omega$), the quadratic limb darkening (LD) coefficients following the \cite{Kipping2013} parametrization, the scaled planetary radius, $R_{P}/R_{\star}$, and the stellar mass, $M_{\star}$, and radius, $R_{\star}$. We assigned Gaussian priors on the stellar mass and radius obtained from the stellar analysis in Section \ref{subsec:thomas}. We also imposed Gaussian priors for the LD coefficients, obtained with the code {\tt PyLDTk}\footnote{Available at \url{https://github.com/hpparvi/ldtk}}\citep{Parviainen2015, Husser2013} for each photometric dataset, but increasing the errors to 0.1, in order to avoid significant deviations between measured and predicted LD coefficients \citep{PatelandEspinoza2022}. The impact parameter, $b$, was set as a free parameter (e.g., \citealt{Frustaglietal2020}) and the dilution factor was not included in the fit since the PDCSAP had already been corrected for crowding (see Sect.~\ref{sec:obs}).

\begin{figure}
\centering
\includegraphics[width=0.99\linewidth]{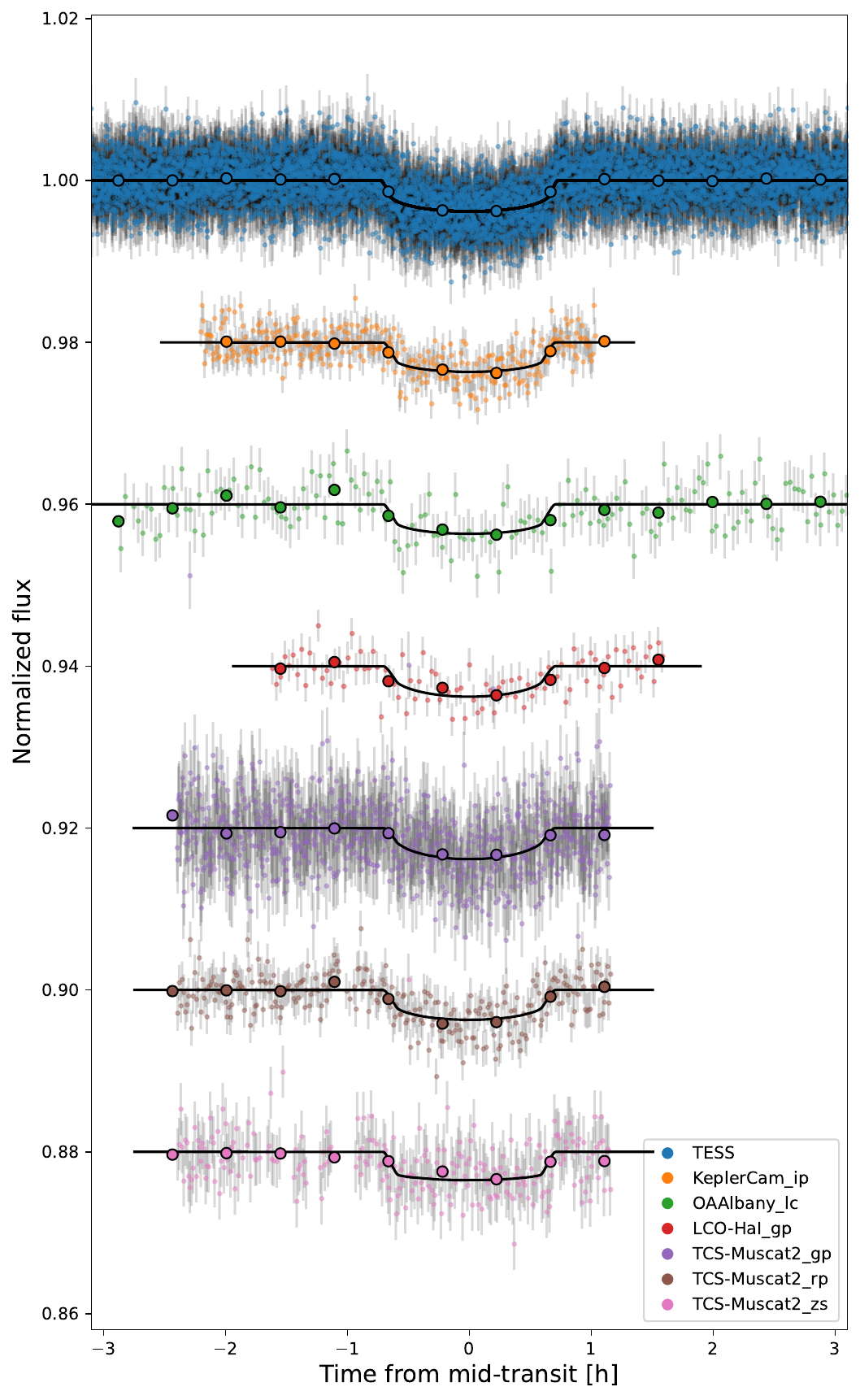}
\caption{TESS and ground-based light curves with binned data for TOI-3862\,b, together with the models obtained from the joint fit.}\label{fig:transits_fit}
\end{figure}

\begin{figure}
\centering
\includegraphics[width=0.99\linewidth]{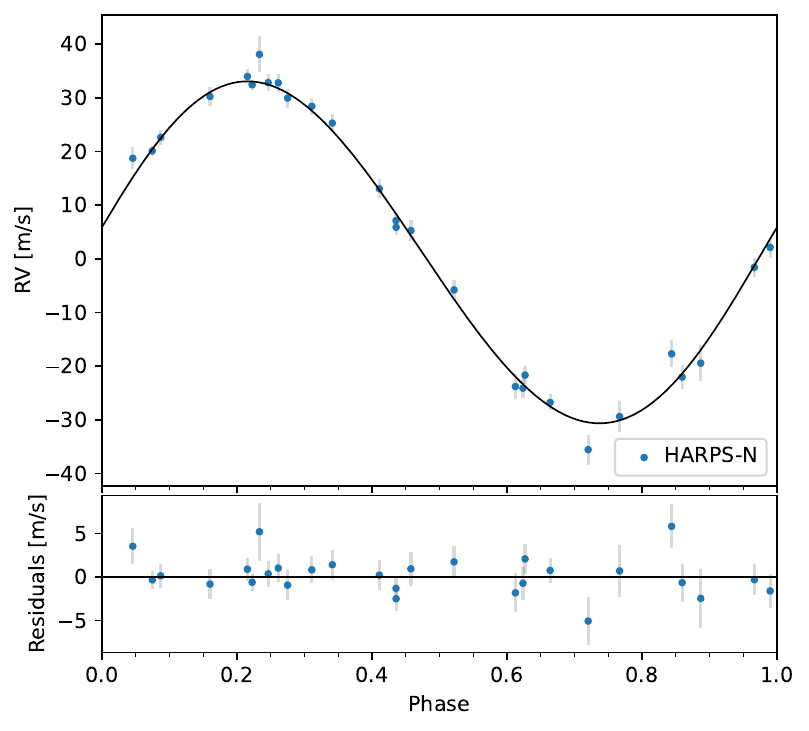}
\caption{HARPS-N  RV data for TOI-3862, with the 1p+GP model overplotted.}\label{fig:rv_fit}
\end{figure}

Possible systematics and stellar activity noise were taken into account in the RV data by adding offset and jitter terms to the fit. The parameter space was sampled by using the dynamic nested sampler  \texttt{dynesty}\footnote{\url{https://github.com/joshspeagle/dynesty}} \citep{Speagle2020, Koposovetal2022}, employing 1000 live points. 
We tested models both with and without a GP component. When including the GP, we utilized the \texttt{george} package \citep{george}, adopting a quasi-periodic kernel as defined by \cite{2015ApJ...808..127G}. In this formulation, $h$ denotes the amplitude of the correlations, $\theta$ corresponds to the stellar rotation period, $\omega$ characterizes the length scale of the periodic component (related to the evolution of active regions), and $\lambda$ represents the decay timescale of the correlations. To assess the significance of the tested models, we calculated the Bayesian evidence (log$\mathcal{Z}$) from the nested sampling procedure. We found a difference of $\Delta$log$\mathcal{Z} > 3$ in favor of the more complex GP model, which is not enough to be considered as the preferred model \citep{jeffreys1961theory}. In addition, the posterior distributions of the model parameters are consistent with each other, so we adopt hereafter the values obtained from the model without GP. We find that TOI-3862\,b has an orbital period of 1.56 days and an eccentricity of 0.0232$_{-0.0054}^{+0.0071}$. The planet has a mass of 53.7$_{-2.9}^{+2.8}$ M$_{\oplus}$ and a radius of $5.53 \pm 0.18$ R$_{\oplus}$. The full set of system parameters derived from the fit is summarized in Table \ref{tab:fit_params_toi3862}, the transit fits are shown in Figure \ref{fig:transits_fit}, and the RV model fit is shown in Figure \ref{fig:rv_fit}. For completeness, we also present the system parameters derived from the fit with the GP in the Appendix \ref{tab:modelGP}.

\begin{table*}
  \footnotesize
  \caption{TOI-3862 parameters from the transit and RV joint fit, obtained with the model without GP, as well as parameters from internal structure retrievals and the atmospheric evolution. \label{tab:fit_params_toi3862}}  
  \centering
  \begin{tabular}{lcc}
  \noalign{\smallskip}
  \hline
  \hline
  \noalign{\smallskip}
  Parameter & Prior$^{(\mathrm{a})}$  & Value$^{(\mathrm{b})}$  \\
  \noalign{\smallskip}
  \hline
  \noalign{\smallskip}
  \multicolumn{3}{l}{\emph{ \bf Model Parameters }} \\
    Orbital period $P_{\mathrm{orb}}$ (days)  & $\mathcal{U}[1.2, 1.8]$   & $ 1.55745774_{-4.2\times10^{-7}}^{+3.6\times10^{-7}}$ \\
      \noalign{\smallskip}
    Transit epoch $T_0$ (BJD - 2,450,000)  & $\mathcal{U}[9662.0, 9663.0]$   & $ 9662.73870_{-0.00014}^{+0.00013}$  \\
      \noalign{\smallskip}
    $\sqrt{e} \sin \omega_\star$ &  $\mathcal{U}(-1,1)$ & $ -0.148_{-0.021}^{+0.019}$ \\
          \noalign{\smallskip}
    $\sqrt{e} \cos \omega_\star$  &  $\mathcal{U}(-1,1)$ & $0.031_{-0.027}^{+0.022}$ \\
          \noalign{\smallskip}
    Scaled planetary radius $R_\mathrm{p}/R_{\star}$ &  $\mathcal{U}[0,0.5]$ & $0.05508_{-0.00033}^{+0.00036}$   \\
    Impact parameter, $b$ &  $\mathcal{U}[0,1]$  & $ 0.135_{-0.058}^{+0.059}$ \\
    Radial velocity semi-amplitude variation $K$ (m s$^{-1}$) &  $\mathcal{U}[0,50]$ & $ 31.77\pm0.24 $  \\ 
          \noalign{\smallskip}
    \hline
          \noalign{\smallskip}
    \multicolumn{3}{l}{\textbf{Derived parameters}} \\
    Planet radius ($R_{\rm J}$)  & $\cdots$ & $ 0.493\pm0.016 $  \\
    Planet radius ($R_{\oplus}$)  & $\cdots$ & $ 5.53\pm0.18 $  \\
    Planet mass ($M_{\rm J}$)  & $\cdots$ & $0.1688_{-0.0090}^{+0.0087}$    \\
    Planet mass ($M_{\oplus}$)  & $\cdots$ & $53.7_{-2.9}^{+2.8}$    \\
    Eccentricity $e$  & $\cdots$ & $0.0232_{-0.0054}^{+0.0071}$   \\
    Scaled semi-major axis $a/R_\star$   & $\cdots$ & $ 8.848_{-0.070}^{+0.074} $  \\
    Semi-major axis $a$ (AU)  & $\cdots$ & $ 0.02539_{-0.00068}^{+0.00064} $ \\
    $\omega_{\rm P} $ (deg)  &  $\cdots$ &  $ 278.72_{-0.50}^{+0.48}$  \\
    Orbital inclination $i$ (deg)  & $\cdots$ & $89.15\pm0.37$ \\
    Transit duration $T_{41}$ (days) & $\cdots$ & $ 0.05871_{-0.00049}^{+0.00050}$ \\
    Transit duration $T_{32}$ (days) & $\cdots$ & $ 0.05243_{-0.00049}^{+0.00051}$ \\
\noalign{\smallskip}
     \hline
\noalign{\smallskip}
    \multicolumn{3}{l}{\emph{\bf Calculated parameters}} \\
    Equilibrium Temperature T$_{eq}$ (K) &  & $1539\pm36$ \\
    Planetary density $\rho_P$ (g cm$^{-3}$) &  & $1.75\pm0.20$   \\
\noalign{\smallskip}
         \hline
\noalign{\smallskip}
    \multicolumn{3}{l}{\emph{\bf Other system parameters}} \\
    Jitter term $\sigma_{\rm HARPS-N}$ (\ms) & $\mathcal{U}[0,60]$ & $0.63_{-0.42}^{+0.64}$   \\
    Limb darkening $q_1, TESS$  & $\mathcal{N}[0.4586,0.1]$ & $0.4486\pm0.0025$   \\
    Limb darkening $q_2, TESS$ & $\mathcal{N}[0.1153,0.1]$ & $0.1035_{-0.0029}^{+0.0025}$   \\
    Limb darkening $q_1, KeplerCam$$^{(\mathrm{c})}$  & $\mathcal{N}[0.4740, 0.1]$ & $0.475_{-0.055}^{+0.038}$   \\
    Limb darkening $q_2, KeplerCam$$^{(\mathrm{c})}$ & $\mathcal{N}[0.1155, 0.1]$ & $0.124_{-0.034}^{+0.048}$   \\
    Limb darkening $q_1, LCO$  & $\mathcal{N}[0.7922, 0.1]$ & $0.782_{-0.047}^{+0.041}$   \\
    Limb darkening $q_2, LCO$ & $\mathcal{N}[0.0185, 0.1]$ & $0.003_{-0.049}^{+0.052}$   \\
    Limb darkening $q_1, Muscat2g$  & $\mathcal{N}[0.7925, 0.1]$ & $0.850_{-0.023}^{+0.040}$   \\
    Limb darkening $q_2, Muscat2g$ & $\mathcal{N}[0.0020, 0.1]$ & $0.032_{-0.032}^{+0.054}$   \\
    Limb darkening $q_1, Muscat2r$  & $\mathcal{N}[0.5699, 0.1]$ & $0.631_{-0.037}^{+0.041}$   \\
    Limb darkening $q_2, Muscat2r$ & $\mathcal{N}[0.0014, 0.1]$ & $0.056_{-0.050}^{+0.039}$   \\
    Limb darkening $q_1, Muscat2z$  & $\mathcal{N}[0.4109, 0.1]$ & $0.384_{-0.046}^{+0.040}$   \\
    Limb darkening $q_2, Muscat2z$ & $\mathcal{N}[0.0009, 0.1]$ & $0.041_{-0.067}^{+0.035}$   \\
\noalign{\smallskip}
         \hline

\noalign{\smallskip}
  \hline
   \noalign{\smallskip}
    \multicolumn{3}{l}{\emph{\bf Results from internal structure retrievals}} \\
    Envelope-to-planet mass fraction (\%) & $\cdots$ & $17.3_{-2.8}^{+2.9}$ \\
    Core and mantle mass fraction (\%) & $\cdots$ & $82.7_{-2.8}^{+2.9}$  \\
    \noalign{\smallskip}
  \hline
   \noalign{\smallskip}
    \multicolumn{3}{l}{\emph{\bf Results from atmospheric evolution}} \\
    Maximum initial mass ($M_{\oplus}$ & $\cdots$ & $70.6_{-2.8}^{+2.4}$ \\
    Initial envelope mass ($M_{\oplus}$) & $\cdots$ & $25.8_{-1.4}^{+2.1}$ \\
    Initial envelope-to-planet mass fraction (\%) & $\cdots$ & $36.5_{-3.1}^{+4.6}$ \\
    Initial radius ($R_{\oplus}$) & $\cdots$ & $8.5_{-0.1}^{+0.2}$ \\
  \end{tabular}
~\\
  \emph{Note} -- $^{(\mathrm{a})}$ $\mathcal{U}[a,b]$ refers to uniform priors between $a$ and $b$, $\mathcal{N}[a,b]$ to Gaussian priors with median $a$ and standard deviation $b$.\\  
  $^{(\mathrm{b})}$ Parameter estimates and corresponding uncertainties are defined as the median and the 16th and 84th percentiles of the posterior distributions.\\
  $^{(\mathrm{c})}$ Same limb-darkening model for KeplerCam and OAAlbany data ($i$ filter).  \\
\end{table*}

\section{Internal structure and atmospheric evolution for TOI-3862\,b}
\label{sec-toi-3862-atm}

We used the \textsc{jade} code\footnote{We used version 1.0.0 of the code, publicly available at \url{https://gitlab.unige.ch/spice_dune/jade}.} to model the internal structure and atmospheric evolution of TOI-3862\,b. This approach constrains the current internal composition of the planet and estimates its mass at the time of disk dispersal.

\textsc{jade} employs a Monte Carlo Markov chain (MCMC) approach to retrieve the planet’s internal structure, including the mantle-to-planet mass fraction, envelope-to-planet mass fraction, and envelope metallicity. The model is constrained by the observed planetary mass, radius (Table \ref{tab:fit_params_toi3862}), and estimated system age (from Sect.~\ref{sec:sed}, since it provides smaller uncertainties while remaining consistent within errors with the other methods), accounting for stellar irradiation (including X-rays and extreme ultra-violet) and internal heating from the planet core.

The planet is modeled with a differentiated structure: an iron core, a silicate mantle, and a hydrogen and/or helium-dominated envelope with a helium mass fraction set to $Y = 0.2$ \citep[for consistency with Neptune;][]{Hubbard1995, Helled2020}. The envelope consists of an upper radiative layer and a lower region where energy is transported by radiation and convection, respectively. The Rosseland mean opacity of the envelope is increased by including trace metals with solar abundances, controlled by the metallicity parameter ($Z_{\rm met}$). The envelope structure is integrated from the top of the atmosphere downward, using a 1D thermodynamic model and polytropic equations of state for the core and mantle \citep{Seager2007}, iterating until the total integrated mass reaches zero at the center \citep[e.g.,][]{Lopez2013, Jin2014}.

We performed three sets of structure retrievals -- at the nominal age of the system (7.5 Gyr) and at the $\pm3\sigma$ age limits -- to explore how age uncertainty affects the inferred structure. Each run uses 30,000 MCMC steps with 30 walkers and 6,000 burn-in steps to ensure convergence in the resulting internal structure's quantities. Across the different assumed ages, the derived quantities vary within their respective 1$\sigma$ uncertainties, so we report the values and uncertainties from the nominal age in Table~\ref{tab:fit_params_toi3862}.

In the next step, we fixed the core and mantle mass fraction, as well as the atmospheric metallicity to a small nonzero value ($Z_{\rm met}$$\sim$0.04) derived from the previous internal structure retrieval. This choice reflects a more realistic atmospheric structure, while the exact close-to-zero value has little influence on the outcome (see \citealt{Attia2025}, Fig.~1).

In contrast to studies in which atmospheric erosion is delayed due to high-eccentricity migration \citep{Attia2021, Attia2025}, allowing the planet to escape strong irradiation at close orbital distance from the young host star, we assume here that the planet migrated early on within the disk and remained near its current close-in orbit for most of its lifetime, experiencing prolonged and intense stellar irradiation. This justifies fixing the planet's orbit in our simulations starting at 10 Myr, as it would have already reached its close-in configuration by then. Since the evolution of the stellar EUV and X-ray flux is accounted for, the resulting mass loss is not necessarily an upper limit, unless the migration occurred later through a different mechanism. This approach allows us to estimate the planet's maximum possible initial mass by fixing it in its current close-in orbit, thereby maximizing its exposure to stellar irradiation and atmospheric erosion over time. The stellar bolometric and XUV (X-ray and extreme ultraviolet) luminosity curves were derived from the GENEC stellar evolution code \citep{Eggenberger2008}, describing how the changing stellar output influences the heating and atmospheric evolution of the planet.

\begin{figure}[!h]
\includegraphics[width=1.00\linewidth,trim=10 60 0 80,clip]{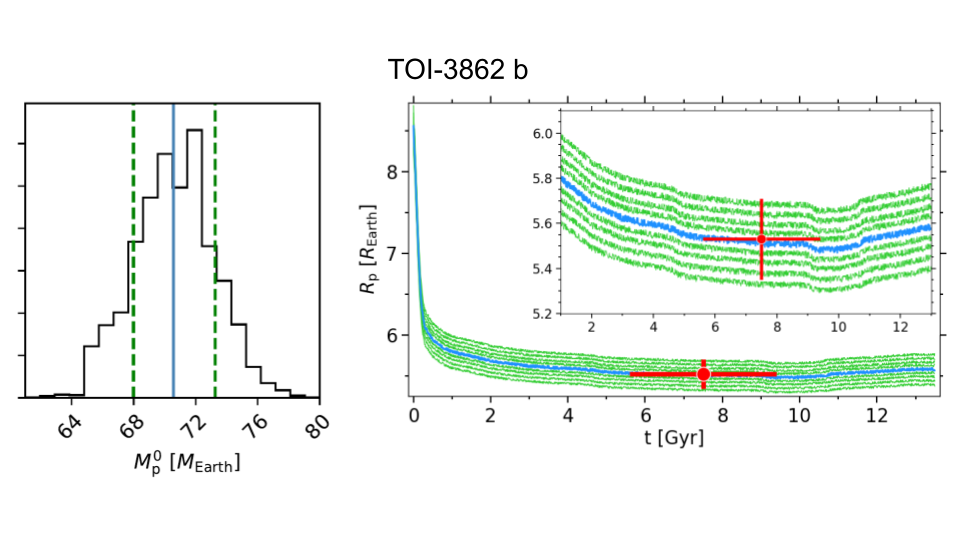}
\caption{\textsc{jade} simulations of atmospheric evolution for TOI-3862\,b. \textit{Left}: Posterior distribution of the initial planet mass. The median is marked by a blue line; the 1$\sigma$ HDI is shown as dashed green lines. \textit{Right}: Temporal evolution of the planet’s radius in the best-fit simulation (blue) and a representative set within the 1$\sigma$ HDI (green). The red point indicates the measured age and radius, with associated uncertainties. An inset shows a zoomed-in view of the planet radius over a narrower range.}  
\label{fig:atm-mass}
\end{figure}

In an early disk-driven migration scenario, we turned off dynamical evolution in the \textsc{jade} simulations and fixed the planet’s orbit to its present-day configuration. The atmospheric evolutionary was run from the expected time of disk dissipation to approximately 13~Gyrs, ensuring full coverage of the systems' age estimates within $3\sigma$. 
The simulation traces the evolution of the planet envelope mass and radius, allowing us to constrain the maximum initial mass of the planet by initializing a suite of simulations from the measured present-day value (0.17~$M_{\rm J}$) up to 1 $M_{\rm J}$. We fit the planetary radius simulated at the estimated system age to the measured present-day value. The retrieval was performed through importance sampling, allowing us to build a posterior distribution for the initial mass and use its median and highest-density intervals (HDIs) as a best-fit value and associated uncertainty range (Figure~\ref{fig:atm-mass}, left panel). This best-fit value, which reproduces the present-day properties through atmospheric escape in a fixed-orbit, high-irradiation scenario, can be interpreted as the planet’s maximum allowable initial mass. Any higher mass would retain too much atmosphere and exceed the observed mass and is therefore excluded.

Figure~\ref{fig:atm-mass} presents the atmospheric evolution of TOI-3862\,b. We performed simulations over a grid of initial planet mass from 0.17 to 1.0~$M_J$. The best-fit simulation (blue curve in right panel) corresponds to an initial mass of $70.6^{+2.7}_{-2.6}~M_{\oplus}$ (or $\sim$0.22~$M_J$) and initial radius of 8.49 R$_{\rm \oplus}$, with an initial envelope mass of $25.8~M_{\oplus}$, corresponding to around $37$\% of the total mass.  Due to TOI-3862\,b's old age ($\sim$7.5~Gyr), it has undergone prolonged atmospheric loss and reached a plateau in its mass and radius evolution. The change in the planet's radius and nature, from a Saturn-size to a Neptune-size planet, is also illustrated in Figure \ref{fig:density_period_diagram_amedeo} (left panel).

\section{Mass-radius relation and hot Neptune desert}
\label{sec:mass-radius}
TOI-3862\,b presents a compelling case study of planetary structure and atmospheric evolution within the hot Neptune desert. With a precisely (19-$\sigma$) measured mass of $53.7^{+2.8}_{-2.9}\,M_\oplus$ and a radius of $5.53 \pm 0.18\,R_\oplus$, TOI-3862\,b lies above the pure water composition curve and below the cold hydrogen model in the mass–radius diagram (see Figure \ref{fig:massradius}). This location is indicative of a planet with a high heavy-element content and a relatively small H/He envelope. These inferences are supported by the internal structure modeling performed with the \textsc{jade} code (see Sect.~\ref{sec-toi-3862-atm}), which suggests that TOI-3862\,b currently retains an H/He envelope comprising $\sim$17\% of the planet's mass, overlying a dense core composed primarily of rock and metal. 

\begin{figure*}         
    \includegraphics[width=0.45\textwidth]{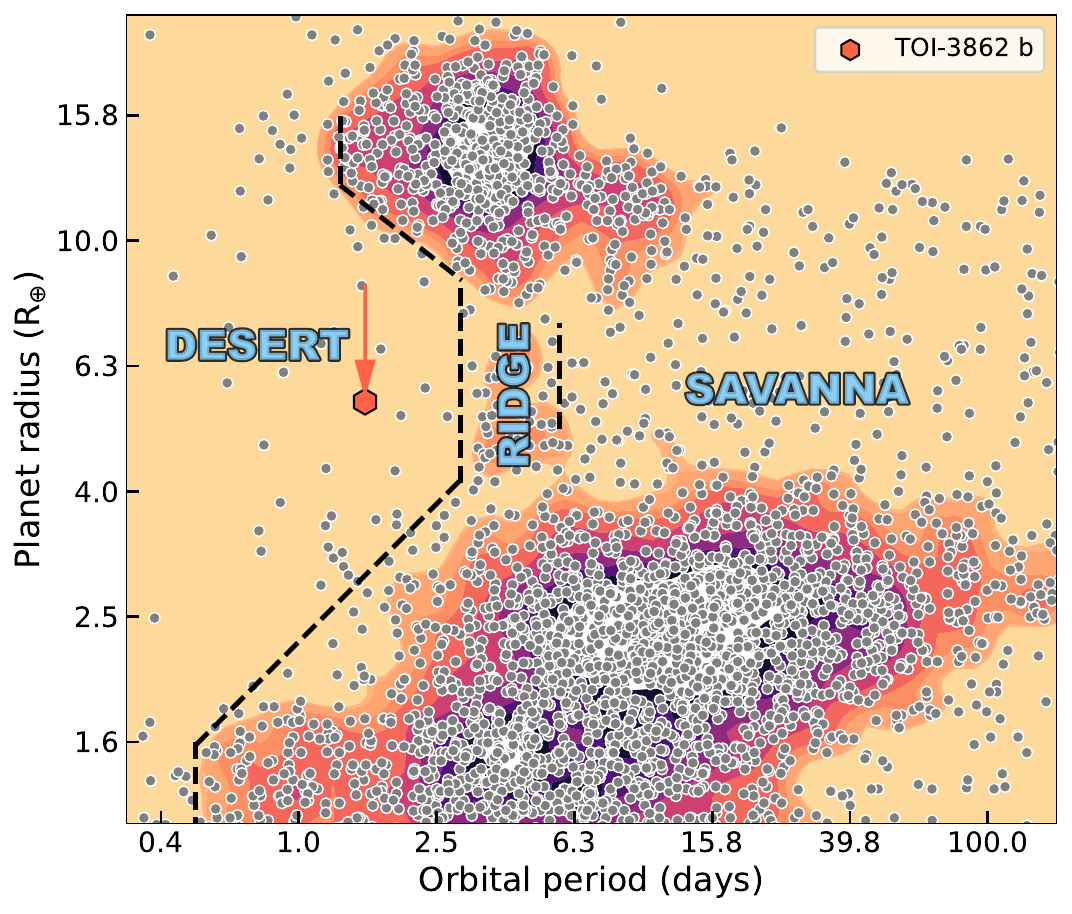}
    \includegraphics[width=0.45\textwidth]{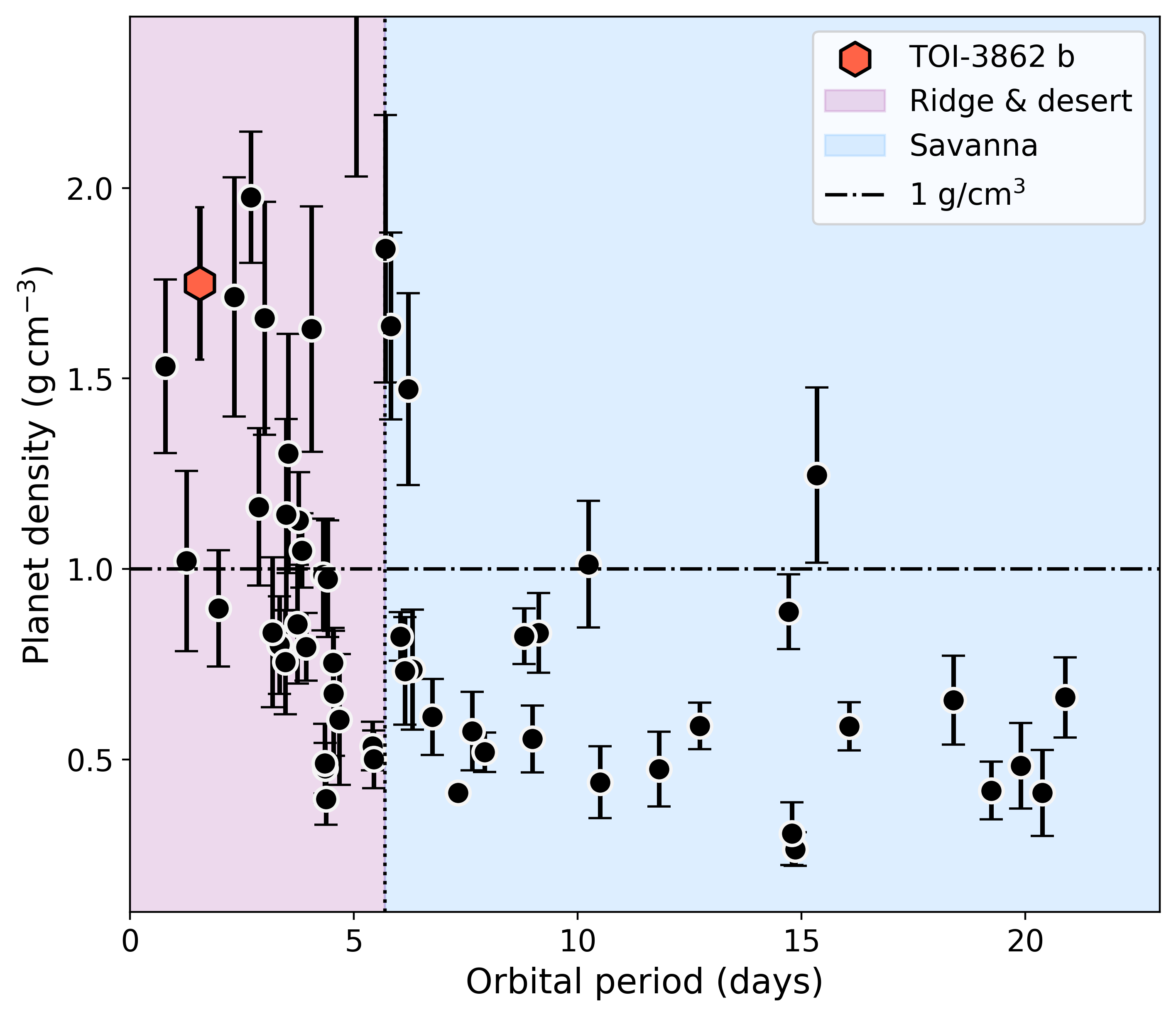}
    \caption{\textit{Left:} Planet radius vs. orbital period for all known exoplanets, highlighting the hot Neptune desert, ridge, and savanna derived by \citealt{CastroGonzalez2024}. We note that this updated definition of the hot Neptune desert is much stricter than the original one by \citealt{Mazehetal2016}, with less than 10\% as many planets in it. The arrows represent the evolution of the planet' radius from the initial value to the current value. \textit{Right:} Density-period diagram of all planets with radii between 4.5 and 8.5 R$_{\rm \oplus}$ and densities constrained to precisions better than 33\%, with the hot Neptune desert planet, TOI-3862\,b, overplotted. These plots were generated with \texttt{nep-des} (\url{https://github.com/castro-gzlz/nep-des}).}
    \label{fig:density_period_diagram_amedeo}
\end{figure*}

\begin{figure}         
    \includegraphics[width=0.49\textwidth]{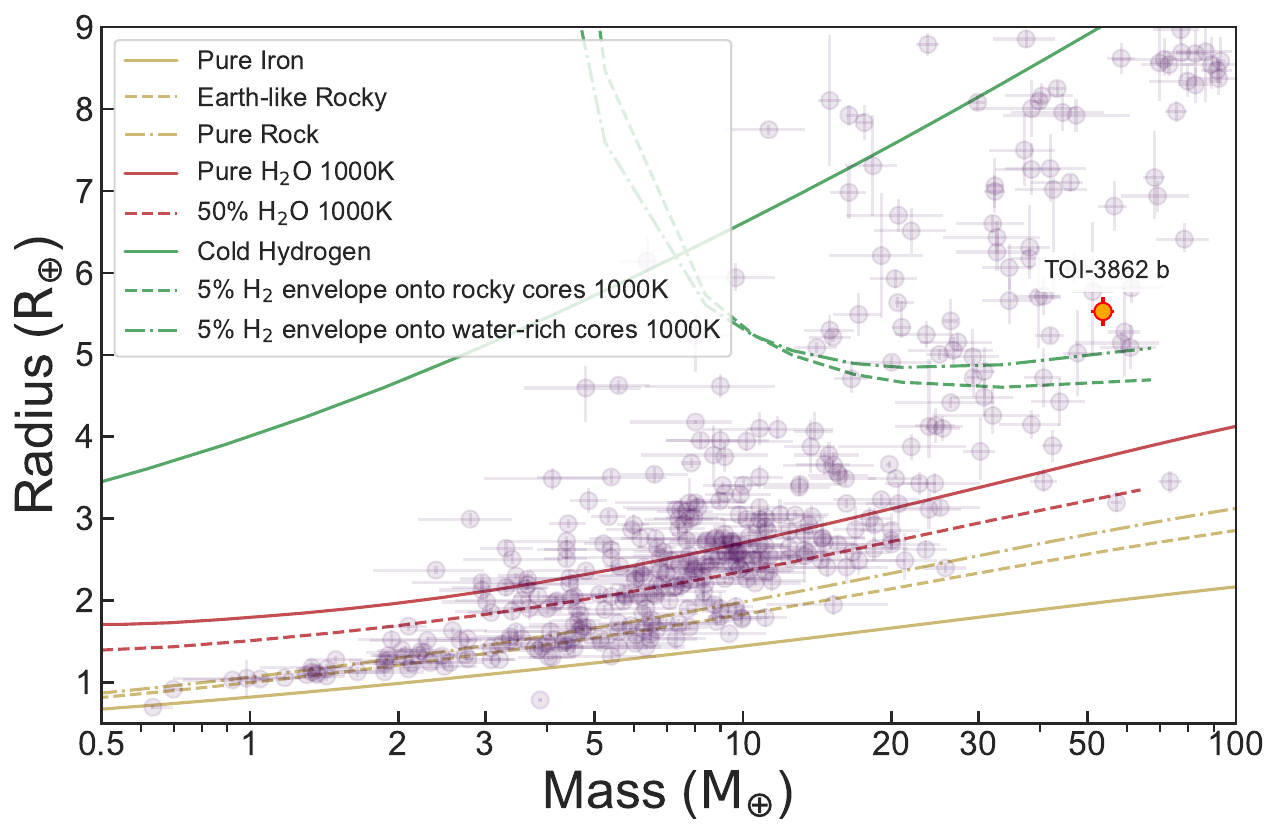}
    \caption{Mass–radius diagram for exoplanets with measured masses and radii, showing the population of known planets alongside theoretical composition models from \citet{Zeng2016} (solid and dashed curves). TOI-3862\,b is overplotted in orange color.}
    \label{fig:massradius}
\end{figure}

With an orbital period of $\sim$1.56 days and a radius of $\sim$5.53~$R_\oplus$, TOI-3862\,b lies squarely in the middle of the so-called hot Neptune desert (see Figure~\ref{fig:density_period_diagram_amedeo}, left panel) — a region of parameter space sparsely populated by planets due to the expected effects of photoevaporation and atmospheric stripping. The existence of planets in this desert poses important constraints on atmospheric retention and migration history, especially in high-irradiation environments. In terms of bulk density, TOI-3862\,b has a measured density of $1.75 \pm 0.20$~g\,cm$^{-3}$, placing it among the densest super-Neptunes in this regime. In particular, TOI-3862\,b follows the empirical density–period trend described by \citealt{CastroGonzalez2024b} (Figure  \ref{fig:density_period_diagram_amedeo}, right panel), where super-Neptune-size planets that survive in the desert tend to have bulk densities exceeding 1 g cm$^{-3}$, similar to the super-Neptunes in the ridge, and in contrast to the longer-period savanna planets, which typically show densities below 1 g cm$^{-3}$. Using the NASA Exoplanet Archive (NEA) catalog\footnote{\url{https://exoplanetarchive.ipac.caltech.edu/}}, we constructed a comparative sample by selecting confirmed exoplanets with orbital periods $\leq$ 3.2 days and radii between 4.5 and 8.5 $R_\oplus$, which defines the empirical boundaries of the desert in the super-Neptune domain \citep{CastroGonzalez2024}. Within this sample, TOI-3862\,b ranks as the second densest planet (after TOI-1288\,b, with mean density of $\sim$1.97~g\,cm$^{-3}$, \citealt{Knudstrupetal2023, Polanskietal2024}), reinforcing the view that intermediate-size desert survivors are rare, dense, and likely possess high heavy-element fractions.

 \section{Conclusions} \label{sec:concl}

We report the confirmation and detailed characterization of TOI-3862\,b, a dense sub-Saturn orbiting a G-type host star with an effective temperature of $\sim$5300 K. TOI-3862\,b has a mass of 53.7$_{-2.9}^{+2.8}$  M$_{\rm \oplus}$, a radius of 5.53 $\pm$ 0.18 R$_{\rm \oplus}$, and an orbital period of $\sim$1.56 days, placing it well within the so-called hot Neptune desert — a region of parameter space sparsely populated due to efficient atmospheric stripping under intense stellar irradiation. From its measured bulk properties, we determine a density of 1.75\,$\pm$\,0.20~g\,cm$^{-3}$. On the mass–radius diagram, TOI-3862\,b lies above the pure water curve and below the cold hydrogen model, consistent with a metal-rich composition and a relatively modest gaseous envelope. This is corroborated by internal structure modeling using the \textsc{jade} code: despite its strong irradiation environment, our modeling suggests that the planet retains a minimal H/He envelope (15–17\% H/He envelope, 40–45\% silicate mantle, and 38–41\% iron core, depending on the assumed system age), with nearly all of its original atmosphere having been lost over time. Its radius evolution history supports theories of atmospheric stripping over gigayear timescales. 
In-transit RVs offer a unique opportunity for future studies, particularly for constraining mass loss through high-resolution transmission spectroscopy targeting escaping hydrogen and helium. Such observations could help place limits on atmospheric escape rates in highly irradiated planets. Additionally, in-transit RVs are subject to the Rossiter–McLaughlin (RM) effect, with TOI-3862\,b expected to produce an RM amplitude of approximately 6\,m\,s$^{-1}$. Dedicated RM observations could help characterize the system's spin–orbit alignment, providing valuable insights into its dynamical history and formation pathways.

We further find that TOI-3862\,b follows the empirical density–period trend established by \cite{CastroGonzalez2024b}, whereby a fraction of super-Neptunes in the desert and ridge are found with high densities ($\rho_P$ $\simeq$ 1.5-2.0 g cm$^{-3}$), in contrast to the longer-period low-density ($\rho_P$ $\leq$ 1 g cm$^{-3}$) savanna planets. Within a sample of exoplanets from the NEA catalog with orbital periods of $\leqslant$ 3.2 days and radii of between 4.5 and 8.5~R$_{\rm \oplus}$ \citep[i.e. the desert boundaries for intermediate-size planets;][]{CastroGonzalez2024}, TOI-3862\,b emerges as the second densest known planet in this regime (after TOI-1288\,b), highlighting its significance as a benchmark object for studies of planetary structure and atmospheric evolution under extreme stellar irradiation conditions.

\begin{acknowledgements}
I.C. acknowledge financial contribution from the INAF Large Grant 2023 ``EXODEMO''.
C.M.P., gratefully acknowledge the support of the  Swedish National Space Agency (DNR 65/19). 
We acknowledge financial support from the Agencia Estatal de Investigaci\'on of the Ministerio de Ciencia e Innovaci\'on MCIN/AEI/10.13039/501100011033 and the ERDF “A way of making Europe” through project PID2021-125627OB-C32, and from the Centre of Excellence “Severo Ochoa” award to the Instituto de Astrofisica de Canarias.
T.M. acknowledges financial support from the Spanish Ministry of Science and Innovation (MICINN) through the Spanish State Research Agency, under the Severo Ochoa Program 2020-2023 (CEX2019-000920-S) and from the Spanish Ministry of Science and Innovation with the grant no. PID2023-146453NB-100 (\textit{PLAtoSOnG}). H.J.D acknowledges funding from the same source under grants PID2019-107061GB-C66 and PID2023-149439NB-C41. 
DRC acknowledges partial support from NASA Grant $18-2XRP18_2-0007$. This research has made use of the Exoplanet Follow-up Observation Program (ExoFOP; DOI: $10.26134/ExoFOP5$) website, which is operated by the California Institute of Technology, under contract with the National Aeronautics and Space Administration under the Exoplanet Exploration Program. Based on observations obtained at the Hale Telescope, Palomar Observatory, as part of a collaborative agreement between the Caltech Optical Observatories and the Jet Propulsion Laboratory operated by Caltech for NASA. The Observatory was made possible by the generous financial support of the W. M. Keck Foundation. The authors wish to recognize and acknowledge the very significant cultural role and reverence that the summit of Maunakea has always had within the Native Hawaiian community. We are most fortunate to have the opportunity to conduct observations from this mountain.
G.N. thanks for the research funding from the Ministry of Science and Higher Education programme the "Excellence Initiative - Research University" conducted at the Centre of Excellence in Astrophysics and Astrochemistry of the Nicolaus Copernicus University in Toru\'n, Poland.
D.J., K.G. and G.N. gratefully acknowledges the Centre of Informatics Tricity Academic Supercomputer and networK (CI TASK, Gda\'nsk, Poland) for computing resources (grant no. PT01016).
Some of the observations in this paper made use of the High-Resolution Imaging instrument ‘Alopeke and were obtained under Gemini LLP Proposal Number: GN/S-2021A-LP-105. ‘Alopeke was funded by the NASA Exoplanet Exploration Program and built at the NASA Ames Research Center by Steve B. Howell, Nic Scott, Elliott P. Horch, and Emmett Quigley. Alopeke was mounted on the Gemini North telescope of the international Gemini Observatory, a program of NSF’s OIR Lab, which is managed by the Association of Universities for Research in Astronomy (AURA) under a cooperative agreement with the National Science Foundation. on behalf of the Gemini partnership: the National Science Foundation (United States), National Research Council (Canada), Agencia Nacional de Investigación y Desarrollo (Chile), Ministerio de Ciencia, Tecnología e Innovación (Argentina), Ministério da Ciência, Tecnologia, Inovações e Comunicações (Brazil), and Korea Astronomy and Space Science Institute (Republic of Korea). 
J.L.-B. is partially funded by the NextGenerationEU/PRTR grant CNS2023-144309 and national projects PID2019-107061GB-C61 and PID2023-150468NB-I00 by the Spanish Ministry of Science and Innovation/State Agency of Research MCIN/AEI/10.13039/501100011033.
Support for this work was provided by NASA through the NASA Hubble Fellowship grant HST-HF2-51559.001-A awarded by the Space Telescope Science Institute, which is operated by the Association of Universities for Research in Astronomy, Inc., for NASA, under contract NAS5-26555.
This project has received funding from the European Research Council (ERC) under the European Union's Horizon 2020 research and innovation programme (project {\sc Spice Dune}, grant agreement No 947634).
This work has been carried out in the frame of the National Centre for Competence in Research PlanetS supported by the Swiss National Science Foundation (SNSF) under grants 51NF40\_182901 and 51NF40\_205606. P.E. acknowledges support from the
SNF grant No 219745.

This work makes use of observations from the LCOGT network. Part of the LCOGT telescope time was granted by NOIRLab through the Mid-Scale Innovations Program (MSIP). MSIP is funded by NSF.

This paper is based on observations made with the Las Cumbres Observatory’s education network telescopes that were upgraded through generous support from the Gordon and Betty Moore Foundation.

Funding for the TESS mission is provided by NASA's Science Mission Directorate. KAC acknowledges support from the TESS mission via subaward s3449 from MIT.

This work is partly supported by JSPS KAKENHI Grant Number JP24K00689,
JP24K17082, JP24H00017 and JP24H00248, JSPS Grant-in-Aid for JSPS
Fellows Grant Number JP25KJ0091, and JSPS Bilateral Program Number
JPJSBP120249910.
This article is based on observations made with the MuSCAT2
instrument, developed by ABC, at Telescopio Carlos Sánchez operated on
the island of Tenerife by the IAC in the Spanish Observatorio del
Teide.

The authors acknowledge support from the Swiss NCCR PlanetS and the Swiss National Science Foundation. This work has been carried out within the framework of the NCCR PlanetS supported by the Swiss National Science Foundation under grants 51NF40182901 and 51NF40205606. J.K. gratefully acknowledges the support of the Swedish National Space Agency (SNSA; DNR 2020-00104) and of the Swiss National Science Foundation under grant number TMSGI2\_211697.

R.L. acknowledges financial support from the Severo Ochoa grant CEX2021-001131-S funded by MCIN/AEI/10.13039/501100011033 and the European Union (ERC, THIRSTEE, 101164189). Views and opinions expressed are however those of the author(s) only and do not necessarily reflect those of the European Union or the European Research Council. Neither the European Union nor the granting authority can be held responsible for them.

G.M. acknowledges financial support from the Severo Ochoa grant CEX2021-001131-S and from the Ramón y Cajal grant RYC2022-037854-I funded by MCIN/AEI/1144 10.13039/501100011033 and FSE+.

This research made use of \texttt{nep-des} (available in \url{https://github.com/castro-gzlz/nep-des})

This work made use of \texttt{TESS-cont} (\url{https://github.com/castro-gzlz/TESS-cont}), which also made use of \texttt{tpfplotter} \citep{aller2020} and \texttt{TESS-PRF} \citep{2022ascl.soft07008B}.

 \end{acknowledgements}

\bibliographystyle{aa}
\bibliography{references}

\begin{appendix}

\section{RV datasets}\label{app:rv_data}
\begin{table}[H]
\caption{\label{tab:rvdata_toi3862} Time series of TOI-3862 from HARPS-N data: Julian dates, RVs and their related uncertainties.}
\begin{tabular}{ccc}
\hline
\noalign{\smallskip}
JD   &      RV  & $\sigma_{\rm RV}$    \\ 
        & (\ms) &    (\ms)    \\
\hline             
\noalign{\smallskip}
 2459664.511340   &       -28183.83   &       3.85     \\
 2459664.679194   &       -28199.66   &       3.99   \\
 2459676.445263   &       -28129.41   &       3.00   \\
 2459677.522290   &       -28163.40   &       2.31  \\
 2459765.470504   &       -28154.19   &       1.75   \\
 2459766.465739   &       -28141.68   &       1.58    \\
 2459785.402480   &       -28128.30   &       5.52   \\
 2459786.420681   &       -28190.92   &       5.98  \\
 2459933.710500   &       -28155.10   &       3.10  \\
 2459952.663262   &       -28185.14   &       2.81 \\
 2460006.533727   &       -28128.47   &       2.04   \\
 2460008.567978   &       -28170.80   &       2.57  \\
 2460013.622346   &       -28191.76   &       5.15 \\
 2460019.629180   &       -28186.81   &       2.69   \\
 2460021.554004   &       -28184.85   &       2.91   \\
 2460023.713715   &       -28145.03   &       1.85 \\
 2460031.524166   &       -28132.79   &       2.24   \\
 2460035.546822   &       -28181.04   &       3.53 \\
 2460037.596621   &       -28133.05   &       2.57 \\
 2460041.496401   &       -28193.09   &       2.35  \\
 2460043.648245   &       -28144.61   &       3.46 \\
 2460064.502789   &       -28154.32   &       2.42 \\
 2460067.423150   &       -28137.69   &       2.72 \\
 2460076.419323   &       -28138.87   &       2.10 \\
 2460106.515482   &       -28148.10   &       2.93  \\
 2460112.451765   &       -28130.05   &       1.57  \\
 2460121.434263   &       -28160.56   &       2.78  \\
 2460134.4411318   &       -28136.39   &       2.51  \\
\noalign{\smallskip}
\hline
\noalign{\smallskip}
\end{tabular}
\end{table}

\begin{table*}[!ht]
\caption{\label{tab:drsdata_toi3862} Time series of TOI-3862 activity indicators from HARPS-N Data Reduction Software: bisector, CCF contrast, CCF FWHM, S-index, \logrhk, and their related uncertainties.}
\begin{tabular}{crccccccccc}
\noalign{\smallskip}
\hline
\noalign{\smallskip}
JD & BIS & $\sigma_{\rm BIS}$ & CONT & $\sigma_{\rm CONT}$ & FWHM & $\sigma_{\rm FWHM}$ & $\rm log\,R^{\prime}_\mathrm{HK}$ & $\sigma_{\rm log\,R^{\prime}_\mathrm{HK}}$ &  S-index  & $\sigma_{\rm S-index}$ \\
 & (\ms) & (\ms) & (\kms) & (\kms) & (\kms) & (\kms) &  &  &  &  \\
 \noalign{\smallskip}
\hline
\noalign{\smallskip}
2459664.511340 & -5.07 & 5.44 & 45.06 & 0.45 & 6.17 & 0.01 & -4.96 & 0.11 & 0.17 & 0.02 \\
2459664.679194 & -0.93 & 5.65 & 45.09 & 0.45 & 6.16 & 0.01 & -5.34 & 0.27 & 0.12 & 0.02 \\
2459676.445263 & -4.34 & 4.24 & 45.13 & 0.45 & 6.15 & 0.01 & -5.12 & 0.12 & 0.14 & 0.02 \\
2459677.522290 & -11.57 & 3.27 & 45.18 & 0.45 & 6.16 & 0.01 & -5.03 & 0.06 & 0.16 & 0.01 \\
2459765.470504 & -6.96 & 2.48 & 45.19 & 0.45 & 6.17 & 0.01 & -5.07 & 0.04 & 0.15 & 0.01 \\
2459766.465739 & -0.54 & 2.24 & 45.21 & 0.45 & 6.16 & 0.01 & -5.02 & 0.03 & 0.16 & 0.01 \\
2459785.402480 & -4.52 & 7.80 & 45.00 & 0.45 & 6.17 & 0.01 & -5.39 & 0.46 & 0.12 & 0.03 \\
2459786.420681 & -13.55 & 8.45 & 45.08 & 0.45 & 6.19 & 0.01 & -6.24 & 4.10 & 0.09 & 0.04 \\
2459933.710500 & -14.94 & 4.38 & 45.11 & 0.45 & 6.18 & 0.01 & -5.17 & 0.13 & 0.14 & 0.02 \\
2459952.663262 & 3.45 & 3.98 & 44.61 & 0.45 & 6.17 & 0.01 & -5.03 & 0.08 & 0.16 & 0.01 \\
2460006.533727 & 2.19 & 2.88 & 45.09 & 0.45 & 6.16 & 0.01 & -5.03 & 0.05 & 0.16 & 0.01 \\
2460008.567978 & -16.59 & 3.63 & 44.79 & 0.45 & 6.16 & 0.01 & -4.93 & 0.06 & 0.18 & 0.01 \\
2460013.622346 & -18.43 & 7.28 & 44.24 & 0.44 & 6.17 & 0.01 & -5.11 & 0.22 & 0.15 & 0.03 \\
2460019.629180 & -4.69 & 3.80 & 45.17 & 0.45 & 6.16 & 0.01 & -4.95 & 0.06 & 0.17 & 0.01 \\
2460021.554004 & -11.58 & 4.11 & 45.14 & 0.45 & 6.16 & 0.01 & -5.06 & 0.09 & 0.15 & 0.01 \\
2460023.713715 & -46.50 & 2.62 & 58.29 & 0.58 & 6.72 & 0.01 & -4.98 & 0.04 & 0.17 & 0.01 \\
2460031.524166 & -5.35 & 3.16 & 45.12 & 0.45 & 6.16 & 0.01 & -4.93 & 0.05 & 0.18 & 0.01 \\
2460035.546822 & 3.60 & 4.99 & 44.53 & 0.45 & 6.16 & 0.01 & -5.00 & 0.10 & 0.16 & 0.02 \\
2460037.596622 & -3.33 & 3.64 & 44.78 & 0.45 & 6.17 & 0.01 & -4.96 & 0.06 & 0.17 & 0.01 \\
2460041.496401 & -16.11 & 3.33 & 44.81 & 0.45 & 6.15 & 0.01 & -5.07 & 0.07 & 0.15 & 0.01 \\
2460043.648244 & -6.08 & 4.90 & 44.33 & 0.44 & 6.16 & 0.01 & -5.01 & 0.10 & 0.16 & 0.02 \\
2460064.502789 & -4.58 & 3.42 & 45.06 & 0.45 & 6.16 & 0.01 & -5.11 & 0.08 & 0.15 & 0.01 \\
2460067.423150 & 3.62 & 3.84 & 44.89 & 0.45 & 6.16 & 0.01 & -4.97 & 0.06 & 0.17 & 0.01 \\
2460076.419323 & -1.98 & 2.97 & 45.18 & 0.45 & 6.17 & 0.01 & -5.04 & 0.05 & 0.16 & 0.01 \\
2460106.515482 & -2.16 & 4.15 & 45.16 & 0.45 & 6.15 & 0.01 & -4.96 & 0.08 & 0.17 & 0.02 \\
2460112.451765 & -8.48 & 2.21 & 45.25 & 0.45 & 6.16 & 0.01 & -5.02 & 0.03 & 0.16 & 0.01 \\
2460121.434263 & -7.76 & 3.94 & 45.14 & 0.45 & 6.15 & 0.01 & -5.06 & 0.09 & 0.15 & 0.01 \\
2460134.441132 & -2.18 & 3.54 & 45.23 & 0.45 & 6.16 & 0.01 & -4.98 & 0.07 & 0.17 & 0.01 \\
\noalign{\smallskip}
\hline
\noalign{\smallskip}
\end{tabular}
\end{table*}


\begin{table*}[!ht]
\caption{\label{tab:servaldata_toi3862} Time series of TOI-3862 activity indicators from \texttt{serval}: chromospheric index CRX, differential line width dLW, H-alpha, the sodium lines Na$_1$ and Na$_2$, and their related uncertainties.}
\begin{tabular}{crcrccccccc}
\noalign{\smallskip}
\hline
\noalign{\smallskip}
JD & CRX & $\sigma_{\rm CRX}$ & DLW & $\sigma_{\rm DLW}$ & H$\alpha$ & $\sigma_{\rm H\alpha}$ & Na1 & $\sigma_{\rm Na1}$ & Na2 & $\sigma_{\rm Na2}$ \\
\noalign{\smallskip}
\hline
\noalign{\smallskip}
2459664.511340 & -29.18 & 18.41 & -32.28 & 3.23 & 0.47 & 0.01 & 0.23 & 0.01 & 0.31 & 0.01 \\
2459664.679194 & -13.59 & 22.83 & -34.05 & 3.74 & 0.48 & 0.01 & 0.23 & 0.01 & 0.31 & 0.01 \\
2459676.445263 & -4.20 & 14.97 & -38.12 & 3.00 & 0.47 & 0.01 & 0.24 & 0.01 & 0.33 & 0.01 \\
2459677.522290 & 8.46 & 14.37 & -43.15 & 2.79 & 0.47 & 0.01 & 0.24 & 0.01 & 0.32 & 0.01 \\
2459765.470504 & 5.55 & 11.54 & -42.37 & 2.02 & 0.49 & 0.01 & 0.24 & 0.01 & 0.32 & 0.01 \\
2459766.465739 & 21.28 & 8.30 & -43.73 & 1.71 & 0.49 & 0.01 & 0.24 & 0.01 & 0.31 & 0.01 \\
2459785.402480 & -22.53 & 26.97 & -35.37 & 5.43 & 0.50 & 0.01 & 0.26 & 0.01 & 0.34 & 0.01 \\
2459786.420681 & 23.49 & 27.48 & -27.47 & 5.06 & 0.50 & 0.01 & 0.27 & 0.01 & 0.35 & 0.01 \\
2459933.710500 & 1.26 & 16.29 & -34.50 & 3.01 & 0.48 & 0.01 & 0.24 & 0.01 & 0.32 & 0.01 \\
2459952.663262 & 13.67 & 14.32 & -13.31 & 3.51 & 0.48 & 0.01 & 0.23 & 0.01 & 0.32 & 0.01 \\
2460006.533727 & 24.65 & 10.29 & -35.97 & 2.14 & 0.48 & 0.01 & 0.24 & 0.01 & 0.32 & 0.01 \\
2460008.567978 & 0.50 & 15.13 & -20.67 & 2.84 & 0.48 & 0.01 & 0.24 & 0.01 & 0.33 & 0.01 \\
2460013.622346 & 12.14 & 24.59 & 8.27 & 4.55 & 0.48 & 0.01 & 0.24 & 0.01 & 0.32 & 0.01 \\
2460019.629180 & -7.73 & 15.34 & -36.57 & 2.88 & 0.49 & 0.01 & 0.23 & 0.01 & 0.31 & 0.01 \\
2460021.554004 & 12.63 & 18.67 & -39.77 & 2.87 & 0.48 & 0.01 & 0.23 & 0.01 & 0.32 & 0.01 \\
2460023.713715 & -13.33 & 11.20 & -46.05 & 2.45 & 0.48 & 0.01 & 0.23 & 0.01 & 0.31 & 0.01 \\
2460031.524166 & 13.64 & 13.34 & -35.53 & 2.61 & 0.49 & 0.01 & 0.24 & 0.01 & 0.32 & 0.01 \\
2460035.546822 & 15.42 & 20.44 & -8.22 & 4.06 & 0.49 & 0.01 & 0.24 & 0.01 & 0.32 & 0.01 \\
2460037.596622 & -24.34 & 13.43 & -22.84 & 3.16 & 0.48 & 0.01 & 0.23 & 0.01 & 0.32 & 0.01 \\
2460041.496401 & -7.64 & 12.24 & -19.78 & 2.53 & 0.48 & 0.01 & 0.24 & 0.01 & 0.32 & 0.01 \\
2460043.648244 & -45.63 & 14.67 & -3.61 & 4.10 & 0.49 & 0.01 & 0.25 & 0.01 & 0.33 & 0.01 \\
2460064.502789 & 6.96 & 11.52 & -30.58 & 2.48 & 0.48 & 0.01 & 0.25 & 0.01 & 0.33 & 0.01 \\
2460067.423150 & -13.18 & 12.20 & -34.34 & 2.57 & 0.49 & 0.01 & 0.25 & 0.01 & 0.34 & 0.01 \\
2460076.419323 & 1.78 & 11.07 & -40.99 & 2.44 & 0.48 & 0.01 & 0.25 & 0.01 & 0.33 & 0.01 \\
2460106.515482 & 10.26 & 14.28 & -43.58 & 2.64 & 0.48 & 0.01 & 0.24 & 0.01 & 0.34 & 0.01 \\
2460112.451765 & -10.75 & 7.32 & -46.90 & 1.91 & 0.48 & 0.01 & 0.23 & 0.01 & 0.32 & 0.01 \\
2460121.434263 & -22.41 & 15.38 & -37.98 & 2.92 & 0.48 & 0.01 & 0.24 & 0.01 & 0.32 & 0.01 \\
2460134.441132 & -7.40 & 13.31 & -44.35 & 2.75 & 0.48 & 0.01 & 0.24 & 0.01 & 0.33 & 0.01 \\
\noalign{\smallskip}
\hline
\noalign{\smallskip}

\end{tabular}
\end{table*}

\twocolumn

\section{System Parameters from the fit including the GP} \label{tab:modelGP}

\begin{table}[H]
  \footnotesize
  \caption{TOI-3862 parameters from the transit and RV joint fit, obtained with the model with GP. \label{tab:fit_params_toi3862_GP}}  
  \centering
  \begin{tabular}{lcc}
  \noalign{\smallskip}
  \hline
  \hline
  \noalign{\smallskip}
  Parameter & Prior$^{(\mathrm{a})}$  & Value$^{(\mathrm{b})}$  \\
  \noalign{\smallskip}
  \hline
  \noalign{\smallskip}
  \multicolumn{3}{l}{\emph{ \bf Model Parameters }} \\
    Orbital period $P_{\mathrm{orb}}$ (days)  & $\mathcal{U}[1.2, 1.8]$   & $ 1.55745798_{-5.8\times10^{-7}}^{+5.2\times10^{-7}}$ \\
      \noalign{\smallskip}
    Transit epoch $T_0$ (BJD - 2,450,000)  & $\mathcal{U}[9662.0, 9663.0]$   & $ 9662.73864_{-0.00015}^{+0.00014}$  \\
      \noalign{\smallskip}
    $\sqrt{e} \sin \omega_\star$ &  $\mathcal{U}(-1,1)$ & $ -0.155_{-0.016}^{+0.019}$ \\
          \noalign{\smallskip}
    $\sqrt{e} \cos \omega_\star$  &  $\mathcal{U}(-1,1)$ & $0.033_{-0.021}^{+0.020}$ \\
          \noalign{\smallskip}
    Scaled planetary radius $R_\mathrm{p}/R_{\star}$ &  $\mathcal{U}[0,0.5]$ & $0.05461_{-0.00030}^{+0.00037}$   \\
    Impact parameter, $b$ &  $\mathcal{U}[0,1]$  & $ 0.084_{-0.057}^{+0.051}$ \\
    Radial velocity semi-amplitude variation $K$ (m s$^{-1}$) &  $\mathcal{U}[0,50]$ & $ 31.81_{-0.25}^{+0.22} $  \\ 
          \noalign{\smallskip}
    \hline
          \noalign{\smallskip}
    \multicolumn{3}{l}{\textbf{Derived parameters}} \\
    Planet radius ($R_{\rm J}$)  & $\cdots$ & $ 0.489\pm0.016 $  \\
    Planet radius ($R_{\oplus}$)  & $\cdots$ & $ 5.48\pm0.18 $  \\
    Planet mass ($M_{\rm J}$)  & $\cdots$ & $0.1690_{-0.0090}^{+0.0087}$    \\
    Planet mass ($M_{\oplus}$)  & $\cdots$ & $53.7_{-2.7}^{+2.6}$    \\
    Eccentricity $e$  & $\cdots$ & $0.0256_{-0.0057}^{+0.0054}$   \\
    Scaled semi-major axis $a/R_\star$   & $\cdots$ & $ 8.885_{-0.056}^{+0.059} $  \\
    Semi-major axis $a$ (AU)  & $\cdots$ & $ 0.02539_{-0.00068}^{+0.00064} $ \\
    $\omega_{\rm P} $ (deg)  &  $\cdots$ &  $ 278.67_{-0.41}^{+0.38}$  \\
    Orbital inclination $i$ (deg)  & $\cdots$ & $89.47_{-0.32}^{+0.36}$ \\
    Transit duration $T_{41}$ (days) & $\cdots$ & $ 0.05874_{-0.00046}^{+0.00043}$ \\
    Transit duration $T_{32}$ (days) & $\cdots$ & $ 0.05259_{-0.00045}^{+0.00041}$ \\
\noalign{\smallskip}
     \hline
\noalign{\smallskip}
    \multicolumn{3}{l}{\emph{\bf Stellar activity GP model Parameters}} \\
    $h_{\rm HARPS-N}$  (\ms)  &  $\mathcal{U}[0, 100]$ & $0.32\pm0.20$  \\
    $\lambda$  (days)  &  $\mathcal{U}[5, 2000]$& $ 866_{-488}^{+644}$ \\
    $\omega$    &  $\mathcal{U}[0.01, 0.60]$ &$0.098_{-0.072}^{+0.190}$ \\
    $\theta$ (P$_{\rm rot}$)  (days)  &  $\mathcal{U}[2, 100]$ & $33.4 _{-3.5}^{+3.7}$ \\
\noalign{\smallskip}
  \hline
  \end{tabular}
~\\
  \emph{Note} -- $^{(\mathrm{a})}$ $\mathcal{U}[a,b]$ refers to uniform priors between $a$ and $b$, $\mathcal{N}[a,b]$ to Gaussian priors with median $a$ and standard deviation $b$.\\  
  $^{(\mathrm{b})}$ Parameter estimates and corresponding uncertainties are defined as the median and the 16th and 84th percentiles of the posterior distributions.\\
\end{table}

\clearpage

\end{appendix}
 
\end{document}